\documentclass[11pt]{article}
\pdfoutput=1
\usepackage[utf8]{inputenc}
\usepackage{multirow}
\usepackage{amsmath, amsfonts, amssymb}
\usepackage{array}
    \newcolumntype{C}[1]{>{\centering\arraybackslash$}p{#1}<{$}}
\usepackage{comment}
\usepackage{graphicx}
\usepackage{cite}
\usepackage{pdflscape}
\usepackage{psfrag}
\usepackage{amsthm}
\usepackage{enumerate}
\usepackage{arydshln}
\usepackage{pifont}
    
\usepackage{soul}
\usepackage{slashed}
\usepackage{subcaption}
\usepackage{mathrsfs}
\usepackage{ytableau}
 \usepackage{a4wide}
  \usepackage{tikz}
  \usepackage{tikz-cd}
  \usetikzlibrary{shapes.geometric}
  \usepackage{tcolorbox}
  \usepackage[table]{xcolor}
  \usepackage{colortbl}
  \definecolor{dark-gray}{gray}{0.20}
  \definecolor{gray}{gray}{0.30}
  \definecolor{light-gray}{gray}{0.80}
  \definecolor{dark-red}{rgb}{0.7,0,0}
  \definecolor{dark-green}{rgb}{0.1,0.4,0}
  \definecolor{dark-blue}{rgb}{0.3,0.3,0.7}
  \definecolor{light-blue}{rgb}{0.8,0.8,1}
      \definecolor{swamp}{RGB}{240, 199, 197}
      
  \usepackage{pifont}

\usepackage{newunicodechar} 
\usepackage{setspace}

\usepackage{ifthen}
\renewcommand*{\arraystretch}{0.9}
\usepackage{longtable}

\newcommand{\be}{\begin{equation}}
\newcommand{\ee}{\end{equation}}
\newcommand{\eq}[1]{(\ref{#1})}

\def\be{\begin{equation}}
\def\ee{\end{equation}}
\def\bea{\begin{eqnarray}}
\def\eea{\end{eqnarray}}

\newcommand{\Z}{\mathbb{Z}}
\AtBeginDocument{

}

\numberwithin{equation}{section}

\newcommand{\D}[1][\gamma]{\mathbf{D}_{\bf #1}}
\newcommand{\bvec}[1][\gamma]{\vec{b}_{\bf #1}}
\newcommand{\RFM}[2][T]{%
  \ifthenelse{\equal{#1}{T}}%
    {T^{#2}/\Gamma}%
    {\frac{\mathbb{R}^{#2}}{\mathcal{B}}}%
}
\newcommand{\Tr}[2]{{\rm Tr_{\bf #1}}(#2)}

\newcommand{\Vcas}{V_{\text{Cas}}}
\newcommand{\vev}[1]{\langle #1\rangle}
\newcommand{\MPl}{M_\text{Pl}}

\usepackage{jheppub}

\usepackage{cleveref}

\hypersetup{
	colorlinks=true,
	linkcolor=dark-blue,
	citecolor=dark-red,
	urlcolor=dark-green,
	linktoc=page,
    pdftitle={A dS4 maximum in M-theory from Casimir energies on Riemann-flat manifolds},
    pdfauthor={Miquel Aparici, Bruno Valeixo Bento, Miguel Montero},
    pdfkeywords={M-theory,de Sitter,Riemann-flat manifolds,Casimir energies}
}

\theoremstyle{definition}

\theoremstyle{remark}

\crefname{appendix}{Appendix}{Appendices}

\title{A $dS_4$ maximum in M-theory from Casimir energies \\ on Riemann-flat manifolds}

\author{Miquel Aparici$^{1,2}$,} 
\author{Bruno Valeixo Bento$^{2}$} 
\author{and Miguel Montero$^{2,3}$}
\affiliation{$^1$Departamento de F\'{i}sica Te\'{o}rica, Universidad Aut\'{o}noma de Madrid, Cantoblanco, 28049 Madrid, Spain}
\affiliation{$^2$Instituto de F\'{i}sica Te\'{o}rica IFT-UAM/CSIC,
C/ Nicol\'{a}s Cabrera 13-15, Campus de Cantoblanco, 28049 Madrid, Spain}
\affiliation{$^3$CERN, Theoretical Physics Department, 1211 Meyrin, Switzerland}

\emailAdd{miquel.aparici@estudiante.uam.es}
\emailAdd{bruno.bento@ift.csic.es}
\emailAdd{miguel.montero@csic.es}

\abstract{We describe a fully explicit, four-dimensional, top-down de Sitter saddle, constructed via compactification of M-theory on the Riemann-flat manifold $T^7/\mathbb{Z}_8$. The solution has a vacuum energy of $5.02\cdot 10^{-10}\,M_\text{Pl}^4$, a hierarchy $m_{KK}/H\sim 10^{2}$, and is under control with respect to higher loop and higher-derivative corrections. The solution has an internal manifold with a diameter of a few Planck lengths, and it may be unstable due to loops of virtual 11d black hole microstates. While the solution is the first example of a top-down dS maximum from Riemann-flat manifolds in four dimensions and is realised explicitly enough to compute the vacuum energy precisely and the spectrum of light fields, it is far from being phenomenologically viable. The solution also suggests several promising directions in the search for higher quality dS saddles in the Riemann-flat Landscape, which we describe.}

\begin{document}
\emergencystretch 3em
\hypersetup{pageanchor=false}
\makeatletter
\let\old@fpheader\@fpheader
\preprint{\begin{flushright}CERN-TH-2026-227\\ IFT-26-129\end{flushright}}

\makeatother

\maketitle

\section{Introduction and Conclusion}
\label{sec:introduction}

One of the main open questions in String Theory is to construct a concrete top-down solution exhibiting long-lived acceleration. From a phenomenological point of view this challenge is forced upon us by cosmological observations \cite{SupernovaCosmologyProject:1998vns,SupernovaSearchTeam:1998fmf} but, even beyond phenomenology, finding out whether such solutions exist is potentially tied to a greater and deeper understanding of the fundamental constraints imposed by quantum gravity.

Near the boundaries of moduli space, where parametric control can be achieved, accelerating cosmologies are not expected to arise \cite{Dine:1985he,Ooguri:2006in,Lee:2019wij}; in fact, it has been conjectured \cite{Palti:2019pca,vanBeest:2021lhn} within the Swampland programme that acceleration may not exist in asymptotic regions of moduli space \cite{Danielsson:2018ztv,Obied:2018sgi,Ooguri:2018wrx,Andriot:2018wzk,Andriot:2018mav,Bedroya:2019snp}. There is by now ample evidence that this may indeed be the case \cite{Grimm:2019ixq,Calderon-Infante:2022nxb,Etheredge:2024tok}, with the lack of explicit string constructions at the very forefront. Interestingly, it  has also been argued that string theory can harbour eternally accelerating cosmologies \cite{Marconnet:2022fmx,Andriot:2023wvg,Marconnet:2025vhj} provided that no cosmological or apparent horizon is formed \cite{Townsend:2003qv,Hebecker:2018vxz,Marconnet:2022fmx,Andriot:2023wvg,Andriot:2025cyi,Hassfeld:2025hjx}.

The above means that de Sitter backgrounds, which can harbour acceleration, cannot appear in the arbitrarily weakly coupled corners of moduli space. Most constructions proposed so far live at the edge between the asymptotic and strongly coupled regimes: both far enough from the boundaries of moduli space that the asymptotic Swampland conjectures do not kill the construction, but also close enough that a fully non-perturbative formulation of the theory is not required for control (see \cite{Andriot:2026lac} for a recent review). Scenarios such as KKLT \cite{Kachru:2003aw} and LVS \cite{Balasubramanian:2005zx} are found in this very regime and require a mix of ingredients so difficult to control that their validity remains under debate (see e.g. \cite{Gao:2020xqh, Demirtas:2021nlu,Junghans:2022exo,Bena:2022cwb,Gao:2022fdi,Lust:2022lfc,Lust:2022xoq,Hebecker:2022zme,Bena:2022ive,ValeixoBento:2023nbv,McAllister:2024lnt,Kim:2024dnw,Moritz:2025bsi}).
These typically include stringy sources such as orientifolds, non-perturbative effects, and/or higher-derivative corrections, which are beyond the reach of 10 or 11-dimensional supergravity, in order to escape a no-go theorem \cite{Gibbons:1984kp,Maldacena:2000mw} that excludes de Sitter solutions with only classical supergravity ingredients (fluxes and curvature). This necessarily complicates the constructions, for instance raising questions regarding the backreaction of stringy singular sources \cite{Gao:2020xqh}. 

In \cite{ValeixoBento:2025yhz} we opened the door to the systematic exploration of a different region of the Landscape: that of non-supersymmetric Riemann-flat flux compactifications with Casimir energy, setting up the grounds for the explicit computation of Casimir energies on Riemann-flat manifolds (RFMs). We illustrated the potential of this RFM Landscape and the  machinery we constructed by presenting an explicit and well-controlled five-dimensional $dS_5$ maximum solution in M-theory\footnote{There were some algebraic mistakes in the original construction that have been fixed in the latest version of the paper.}. An implementation of the whole machinery is now publicly available as a \emph{Mathematica} package \cite{ValeixoBento:2026jdx} that can be used straightforwardly in the exploration of the RFM Landscape.

Since all ingredients in this setup can be described within supergravity, we are safe from the complications coming from the stringy ingredients of other scenarios. Just as importantly, the validity of any particular solution can be determined algorithmically, via field theory calculations, making the RFM Landscape much easier to explore and any solution much easier to control. We have also established general theorems constraining this Landscape. For instance, there is no dS$_d$ minimum in Casimir-flux compactifications on Riemann-flat manifolds for $d>4$, and $dS_4$ minima are ruled out for all cases starting with a Minkowski theory in either 16 or 32 supercharges \cite{ValeixoBento:2025qih}. This leaves open the possibility of realising dS$_3$ minima or dS$_d$ maxima with $d\geq 4$, both of which remain interesting from both a fundamental and phenomenological perspective. 

In this paper, we report progress on this programme by presenting an example of a well-controlled \emph{four-dimensional} de Sitter solution of 11d supergravity. This solution, which is a four-dimensional analogue of the construction in \cite{ValeixoBento:2025yhz}, is obtained from compactification of M-theory on the Riemann-flat manifold $T^7/\Z_8$; just like in \cite{ValeixoBento:2025yhz}, the maximum  arises from the balance between $G_4$ flux and Casimir energy. To the best of our knowledge, this is the first top-down $dS_4$ maximum solution of string or M-theory, realised explicitly enough to compute the vacuum energy precisely and the spectrum of light fields. 
This contrasts with  \cite{Chen:2025rkb} (see also \cite{Baykara:2026jzs}), where robust arguments for the existence of dS maxima in stringy regimes are provided, but in which the vacuum is estimated to be only around $10^{-3}$ in Planck units, with the amount of control not being enough to reliably compute the vacuum energy or the masses of low-lying fields. 
On the other hand, explicit classical $dS_4$ saddles of type-II supergravity compactified on group manifolds, with intersecting D-branes and orientifold planes, were found in \cite{Andriot:2021rdy,Andriot:2022way,Andriot:2020wpp,Andriot:2022xjh,Andriot:2022yyj,Andriot:2022bnb,Andriot:2024cct}; while these allow for systematic stability analyses, a parametrically controlled embedding into string theory has not yet been established and their validity is still unclear.\footnote{See also \cite{Sterckx:2026dvi} for a new classical de Sitter solution of type IIB supergravity evading the no-go theorems through an S-fold quotient of the internal space; this solution is, however, not scale separated and susceptible to $\alpha'$-corrections.}
Similar issues are present in other proposals of dS maxima within supergravity \cite{Haque:2008jz,Danielsson:2009ff,Chen:2025rkb}, with dS minima constructions being even more affected \cite{Kachru:2003aw,Balasubramanian:2005zx,DeLuca:2021pej}.

Our $dS_4$ maximum has a vacuum energy of $5.02\cdot 10^{-10}$ in four-dimensional Planck units, it is scale separated (with $m_{KK}/H\sim 10^2$), and the moduli have masses of the order of the Hubble scale. The solution appears to be safe from loop and higher-derivative corrections, and even from known non-perturbative effects. Given that the vacuum energy is still far from $10^{-120}\,M_{\rm Pl}^4$, this is not a good solution for phenomenology; yet it is both a proof of principle that such controlled maximum solutions exist in the four-dimensional RFM Landscape and a stepping stone towards a solution that could be compatible with observations. 

Among the lessons learned from this particular solution is the crucial role of the Weyl-rescaling factor in lowering the vacuum energy: for the particular case of M-theory compactified on a seven-dimensional RFM, it gives $V\sim R^{-18}$, where $R$ is roughly the diameter of the manifold. Therefore, if one could arrange a volume modulus of $R\sim 10^{7}$ in eleven-dimensional Planck units in some other example, this would be enough to reach phenomenological values\footnote{We emphasize that this is far from the value of $R$ we obtain here, which is $R\sim 5.16\,\ell_{11}$.}. On the other hand, for this particular class of flux vacua, $R$ is determined by the Casimir coefficient as $R\sim \mathcal{C}^{1/3}$, which leads us to two other crucial points. First, we would like the Casimir coefficient $\mathcal{C}$ to be as large as possible; being a function of the moduli, this value depends strongly on the point in moduli space where the moduli are stabilised. This is painfully clear in our solution, where the moduli are stabilised away from the point that maximises $\mathcal{C}$. Ideally, we should use an RFM that projects out as many moduli as possible, so that the RFM enhancement of Casimir energies is preserved to the fullest. Furthermore, the power of $1/3$ in $R\sim \mathcal{C}^{1/3}$ significantly diminishes the contribution of a large Casimir coefficient to the volume modulus. This power is a direct consequence of the volume dependence of the flux and Casimir terms that generate the maximum: for our solution, $V_{G_4}\sim R^{-1}$ and $\Vcas\sim R^{-4}$, leading to the ratio of $1/3$. The best-case scenario for a four-dimensional solution would arise from balancing the Casimir potential against a flux term that behaves as $V_\text{flux}\sim R^{-3}$. This is not possible in compactifications of M-theory, and strongly encourages us to consider other higher-dimensional theories (with either 16 or 32 supercharges) as a starting point for the construction of RFM vacua. 

Hopefully, these lessons will pave the way towards a better $dS_4$ solution, one that is able to match cosmological observations. We are led to consider more general RFMs, in particular ones that are non-cyclic and even of non-Abelian holonomy, and higher-dimensional theories other than M-theory; this is precisely the focus of our current and future work.

The rest of the note is organised as follows: in Section~\ref{sec:dS4-maximum0}, we review the basics about Riemann-flat manifolds and how to compute their Casimir energy. Section~\ref{sec:dS4-maximum} contains the details of the proposed solution as well as a study of whether it is under control. Finally, Appendix~\ref{app:A} explains how to determine the Hodge norm of $p$-forms on arbitrary RFMs that are fibrations of a torus over a circle. 

\section{Riemann-flat manifolds and Casimir energies}
\label{sec:dS4-maximum0}

In this section we briefly summarise the necessary background on Riemann-flat manifolds and Casimir energies; a fully detailed analysis can be found in \cite{ValeixoBento:2025yhz} (see also \cite{DallAgata:2025jii,Aparici:2025kjj,ValeixoBento:2026jdx}).
All RFMs are a quotient of $\mathbb{R}^k$ with its standard Euclidean metric, by a subgroup $\mathcal{B}\subset ISO(k)$ of the Euclidean group in $k$ dimensions, 
\begin{equation}
    \mathcal{F}_k=\mathbb{R}^k/\mathcal{B} \,,
    \label{rfmdef}
\end{equation} 
whose group action is free, meaning that no $b\in\mathcal{B}$ has fixed points---the quotient is a manifold, rather than an orbifold. 
The quotient \eqref{rfmdef} can be taken in two steps, where we first quotient $\mathbb{R}^k$ by a lattice $\Lambda\subset\mathbb{R}^k$ generating the translation subgroup to define a torus,
\begin{equation} 
    T^k=\mathbb{R}^k/\Lambda \,,
\end{equation}
and then quotient the torus by $\Gamma$, defined as the quotient of $\mathcal{B}$ by the translation subgroup,
\begin{equation} 
    \Gamma\equiv \mathcal{B}/\sim\,, \quad b\sim b' \quad\text{iff}\quad b'\circ b^{-1}\in \Lambda \,.
\end{equation}
In order for the quotient to be well-defined, elements in $\Gamma$ must preserve the lattice $\Lambda$, meaning that $\Gamma$ is a \emph{finite} group of isometries of $T^k$.
The group $\Gamma$ is also called the point group or holonomy group of the RFM;
in fact, RFMs are the only manifolds that have finite holonomy groups \cite{eee57c5f-3af2-3ab0-b3c0-3d30ad3fad43}. 
In practice, representatives of an element $\gamma\in\Gamma$ are given by affine transformations of the form 
\begin{equation}
    \vec{z}\to\D\,\vec{z}+\bvec \,,
    \label{eq:affine-transformation}
\end{equation}
where $\bvec$ is taken modulo translations and $\D$ is orthogonal.
Demanding that the $\Gamma$ action has no fixed points, so that $\mathcal{B}$ is torsion-free, leads to constraints on the pairs $(\mathbf{D}_\gamma,\vec{b}_\gamma)$. For instance, the equation 
\begin{equation} 
    (\mathbf{D}_\gamma-\mathbf{I})\, \vec{z}+\vec{b}_\gamma=0\,\, \text{mod}\,\Lambda 
    \label{fpfcond}
\end{equation}
must have no solution, which requires $\vec{b}_\gamma$ to have a non-zero component along the invariant subspace of $\mathbf{D}_\gamma$ and thus $\mathbf{D}_\gamma$ must have a non-trivial invariant subspace.

The most general metrics on the RFM are those metrics $\mathbf{G}$ on $T^k$ that satisfy \cite{ValeixoBento:2025yhz}
\begin{equation}\label{eq:moduli.fixing}
    \D^T\cdot\mathbf{G}\cdot\D = \mathbf{G} \,, \quad \forall\,\gamma\in\Gamma \,. 
\end{equation} 
Since all elements $\gamma\in\Gamma$ are generated by a given group of generators $\{\mathbf{g}_1,...,\mathbf{g}_\ell\}$, in such a way that the matrices $\D$ are all products of generators, it is sufficient to check condition \eqref{eq:moduli.fixing} for the smaller set of generators. In the particularly simple case of cyclic holonomy $\Gamma = \Z_n$, with generator $\mathbf{g}$ and $\D \in\{\mathbf{I},\D[g],\ldots,\D[g]^{n-1}\}$ for all $\gamma\in\Z_n$, there is only one such condition.

The Casimir energies depend heavily on the spin structure chosen for the RFM. On the covering torus, a spin structure can be specified by the periodicity of spinor fields $\psi(\vec{z})$ under lattice transformations,
\begin{equation} 
    \psi(\vec{z}+\vec{n})=e^{2\pi i\,\vec{h}\cdot\vec{n}}\, \psi(\vec{z}) \,,
    \label{eq:spin-structure}
\end{equation}
and is encoded in a vector $\vec{h}$ whose entries are $0$ along periodic directions and $1/2$ along anti-periodic directions. A spin structure on the torus descends to a well-defined spin structure on the RFM if it is preserved by the action of the group $\Gamma$ defining the quotient; the vector $\vec{h}$ must satisfy the consistency condition \cite{ValeixoBento:2025yhz}
\begin{equation} 
    (\mathbf{I} - \D)^T\, 
        \vec{h}\in\mathbb{Z}^{k} \,.
    \label{eq:spin-structure-condition-1}
\end{equation}
Moreover, for any $\D$ of order $p$, the action of $\D^p$ is a translation along the vector $p\,\bvec$, which must be in the lattice $\mathbb{Z}^k$. On the other hand, in the spin representation the chosen spin lift satisfies $\mathcal{D}^p_{\gamma}=(-1)^{\mathfrak{s}_{\gamma}} \mathbf{I}$, i.e. it is the identity up to a sign. Therefore, the choice of spin structure on the covering $T^k$ along the direction $p\,\bvec$ is correlated with the choice of spin lift, and we must have
\begin{equation} 
    \mathfrak{s}_\gamma\equiv 2\,p\, \vec{h}\cdot \bvec\quad \text{mod}\,\,2\mathbb{Z} \,. 
    \label{eq:spin-structure-condition-2}
\end{equation}
For even order $p$, $\mathfrak{s}_\gamma$ is fixed regardless of the choice of spin lift, constraining the vector $\vec{h}$ directly. For cyclic RFMs, the combined system of equations given by \eq{eq:spin-structure-condition-1} and \eq{eq:spin-structure-condition-2} always has a solution, which correlates with the fact that all cyclic, mapping tori RFMs admit a spin structure \cite{Hiss16012008}.

Upon compactification of a $D$-dimensional field theory on a $k$-dimensional RFM, the lower-dimensional Casimir energy can be expressed as a sum---over the elements of the quotient group and the massless representations $\mathbf{r}$ of the $D$-dimensional theory---of infinite lattice sums, such that each $\gamma\in\Gamma$ gives a definite contribution $\mathcal{E}(\gamma)$ \cite{ValeixoBento:2025yhz},
\begin{equation}
    \Vcas = \sum_{\gamma\in\Gamma}\sum_{\textbf{r}} \Tr{\bf r}{\D}\,\mathcal{E}(\gamma)
    \,,\quad
    \mathcal{E}(\gamma) = -\hat{\delta}_{\vec{h}}\,\frac{\Gamma(s_\gamma)}{2\pi^{s_\gamma}}\cdot
    \frac{\sqrt{G_\parallel}}{\vert\Gamma\vert} \sum_{\vec{\xi}\in\,\Xi_\gamma} \frac{e^{2\pi i \,\Vec{\beta_\gamma}\cdot\vec{\xi}}}{|\vec{\xi} + \bvec^\parallel|^{2s_\gamma}_\parallel} \,.
    \label{eq:Casimir-general}
\end{equation}
In this formula, $s_\gamma = s - \frac{k-k'}{2}$, the lattice $\Xi_\gamma$ and the vector $\vec{\beta}_\gamma$, as well as the projected metric $\mathbf{G}_\parallel$ and inner product $|\cdot|_\parallel$, depend on the element $\gamma\in\Gamma$ and in particular on the $k'$-dimensional subspace that is invariant under $\D$. 
Notice that for the identity element in $\Gamma$, the invariant subspace is the full lattice $\Z^k$ and the sum stays $k$-dimensional.
The factor $\hat{\delta}_{\vec{h}}$ evaluates to $1$ if there exists a solution to the equation 
\begin{equation}
    (\mathbf{I} - \D)^T\,\vec{\eta} = (\mathbf{I} - \D)^T\,\vec{h} \,,
    \label{eq:eta-condition}
\end{equation}
with $\eta\in\Z^k$, and zero otherwise, in which case the sum $\mathcal{E}(\gamma)$ vanishes identically. Interestingly, the factor $\hat{\delta}_{\vec{h}}$ is related to a field theory version \cite{ValeixoBento:2025yhz} of Atkin-Lehner symmetry \cite{atkin1970hecke,Moore:1987ue,Dienes:1990qh}.
 
The identification of $\Xi_\gamma$ with the invariant lattice of $\mathbf{D}_\gamma$ allows for a nice physical interpretation of each contribution $\mathcal{E}(\gamma)$ to the Casimir potential \eq{eq:Casimir-general}. Each $\mathcal{E}(\gamma)$ is given by a lattice sum just like the one that computes Casimir energies on a torus, but localised on the invariant subspace of $\D$. Since the volume element of this subspace is precisely $\sqrt{G_\parallel}$, we find that  $\mathcal{E}(\gamma)$ behaves exactly as the contribution to the potential that would come from an effective $k'$-dimensional ``Casimir brane'' wrapped on the invariant subspace of $\D$ and with a tension given by \eq{eq:Casimir-general}. 

\section{A \texorpdfstring{$dS_4$}{dS4} maximum in M-theory}
\label{sec:dS4-maximum}

We describe a $dS_4\times \mathcal{F}_7$ compactification of M-theory, where $\mathcal{F}_7$ is a Riemann-flat manifold threaded by $G_4$-flux. The four-dimensional scalar potential has only a Casimir and a $G_4$-flux piece,
\begin{equation}
    V^{(4d)} = \Vcas + V_{G_4} \,,
    \label{totpot}
\end{equation}
with the fluxes chosen such that $V^{(4d)}$ has a saddle point in all directions. After presenting the manifold $\mathcal{F}_7$, we study its Casimir potential using the formulae and tools of \cite{ValeixoBento:2025yhz,ValeixoBento:2026jdx}, provide a detailed analysis of the flux potential, and determine the properties of the $dS_4$ maximum. Finally, we will study corrections to the solution, describing in which sense they are small. 

\subsection{The Riemann-flat manifold \texorpdfstring{$\mathcal{F}_7$}{F7}}
\label{subsec:dS4-maximum-RFM}
Consider a $T^7$ parametrised  by coordinates $\vec{z}=(z_1,z_2,z_3,z_4,z_5,z_6,z_7)$, subject to the identifications
\begin{equation}
    z_i\,\sim\,z_i+1 \,.
\end{equation}
We will quotient this torus by the $\mathbb{Z}_8$ action generated by the affine transformation 
\begin{equation}
    \vec{z}\,\rightarrow\iota_{\mathbf{g}}(\vec{z})=\D[g]\,\vec{z}+\bvec[g] \,, 
    \quad\text{with}\quad 
    \D[g]\equiv\left(
        \begin{array}{*2{C{1.1em}}|*4{C{1.1em}}|*1{C{1.1em}}}
            0 & -1 & 0 & 0 & 0 & 0 & 0 \\
            1 & 0 & 0 & 0 & 0 & 0 & 0 \\ \hline
            0 & 0 & 0 & 0 & 0 & -1 & 0 \\
            0 & 0 & 1 & 0 & 0 & 0 & 0 \\
            0 & 0 & 0 & 1 & 0 & 0 & 0 \\
            0 & 0 & 0 & 0 & 1 & 0 & 0 \\ \hline
            0 & 0 & 0 & 0 & 0 & 0 & 1 \\
        \end{array}
        \right)\,,\quad \bvec[g]=\left(\begin{array}{c}0\\0\\0\\0\\0\\0\\ \frac18\end{array}\right) \,.
    \label{Z8-rfmdef}
\end{equation}
This action has no fixed points on $T^7$, since the vector $\bvec[g]$ lies in the invariant subspace of $\D[g]$. The matrix $\D[g]$ is of order 8, and the manifold $\mathcal{F}_7$ is defined as the quotient
\begin{equation}
    \mathcal{F}_7\equiv T^7/\mathbb{Z}_8 \,.
    \label{cqw}
\end{equation}
This particular RFM was identified in \cite{ValeixoBento:2025yhz} as a promising candidate for a $dS_4$ solution that, however, required the analysis of a 3-modulus problem. For this reason it was not studied in detail. Its holonomy generator $\D[g]$ is built from the direct sum of the companion matrices of the cyclotomic polynomials $\Phi_4(x)$ and $\Phi_8(x)$, which gives rise to a finite matrix of order lcm$(4,8)=8$. The $4\times 4$ block that corresponds to the companion matrix of $\Phi_8(x)$ (of order $8$) is precisely what was used for the $T^6$ quotient of \cite{ValeixoBento:2025yhz} that led to the $dS_5$ maximum solution of M-theory. Here we will show that $\mathcal{F}_7$ does indeed allow for a similarly-controlled $dS_4$ solution.

All Riemann-flat metrics on $\mathcal{F}_7$ come from Riemann-flat metrics on the covering $T^7$ that are invariant under $\D[g]$ \cite{ValeixoBento:2025yhz}. The most general possibility is
\begin{align}
    ds^2 &= G_{ij}dz^i\, dz^j \,, \nonumber \\ 
    \mathbf{G} &=R^2\left(
        \begin{array}{ccccccc}
             e^{u-2 w} & 0 & 0 & 0 & 0 & 0 & 0 \\
             0 & e^{u-2 w} & 0 & 0 & 0 & 0 & 0 \\
             0 & 0 & e^{u+w} \cosh (\zeta) & \frac{e^{u+w} \sinh (\zeta)}{\sqrt{2}} & 0 & -\frac{e^{u+w} \sinh (\zeta)}{\sqrt{2}} & 0 \\
             0 & 0 & \frac{e^{u+w} \sinh (\zeta)}{\sqrt{2}} & e^{u+w} \cosh (\zeta) & \frac{e^{u+w} \sinh (\zeta)}{\sqrt{2}} & 0 & 0 \\
             0 & 0 & 0 & \frac{e^{u+w} \sinh (\zeta)}{\sqrt{2}} & e^{u+w} \cosh (\zeta) & \frac{e^{u+w} \sinh (\zeta)}{\sqrt{2}} & 0 \\
             0 & 0 & -\frac{e^{u+w} \sinh (\zeta)}{\sqrt{2}} & 0 & \frac{e^{u+w} \sinh (\zeta)}{\sqrt{2}} & e^{u+w} \cosh (\zeta) & 0 \\
             0 & 0 & 0 & 0 & 0 & 0 & e^{-6 u} \\
        \end{array}
    \right) \,. 
    \label{eq:RFM-metric}
\end{align}
There are four moduli, parametrised here as $\{R,u,w,\zeta\}$; note that 24 of the 28 torus metric moduli have been fixed by the quotient \eq{cqw}. 
The volume of $\mathcal{F}_7$ is 
\begin{equation}
    \text{vol}(\mathcal{F}_7) = \frac18\text{vol}(T^7)= \frac{R^7}{8} \,,
\end{equation}
only depending on $R$ and reduced by a factor of 8 compared to the volume $\text{vol}(T^7) = R^7$ of the covering $T^7$. The remaining three moduli are shape moduli of the RFM, with $w$ and $u$ encoding the fractions of the volume in each block, and $\zeta$ describing the shape of the $4\times 4$ block.

A de Sitter saddle point only requires that the first derivative of the potential vanishes along every direction. We will now identify a subspace of the moduli space, given by a specific value of $\zeta$, where the partial derivative along this direction vanishes automatically due to symmetry arguments (see also \cite{Parameswaran:2024mrc,Chen:2014fqa}). An affine transformation $\vec{z}\,\rightarrow f(\vec{z})=\mathbf{A}\,\vec{z}+\vec{b}$ of the parent $T^7$ will descend to a well-defined diffeomorphism of $\mathcal{F}_7$ if $f$ maps $\mathbb{Z}_8$ orbits into $\mathbb{Z}_8$ orbits, i.e. when $f(\vec{z})$ is in the normalizer of the $\Z_8$ generated by $\iota_\mathbf{g}(\vec{z})$ \cite{ValeixoBento:2025yhz}. In the particular case where $\D[g]\,\vec{b}=\vec{b}$, this requires  
\begin{subequations}
    \begin{align} 
    \mathbf{A}\cdot \D[g]\cdot \mathbf{A}^{-1}&=\D[g]^{n_1} \,, \label{blorp1} \\ 
    \mathbf{A}\,\bvec[g] &= n_1\,\bvec[g] \quad\text{mod}\,\Z^k \,. \label{blorp2} 
\end{align}
\end{subequations}
For $\D[g]$ as in \eq{Z8-rfmdef}, one possibility is 
\begin{equation}
    \mathbf{A}=
    \left(
    \begin{array}{*7{C{1.2em}}}
         1 & 0 & 0 & 0 & 0 & 0 & 0 \\
         0 & 1 & 0 & 0 & 0 & 0 & 0 \\
         0 & 0 & 1 & 0 & 0 & 0 & 0 \\
         0 & 0 & 0 & -1 & 0 & 0 & 0 \\
         0 & 0 & 0 & 0 & 1 & 0 & 0 \\
         0 & 0 & 0 & 0 & 0 & -1 & 0 \\
         0 & 0 & 0 & 0 & 0 & 0 & 1 \\
    \end{array}
    \right),
    \label{az4}
\end{equation}
a matrix of order 2 that satisfies \eqref{blorp1} with $n_1=5$. It does not satisfy the second condition \eqref{blorp2}; however, in the Casimir potential, the twisted sector terms depend only on the metric restricted to the invariant subspace (which never spans the coordinates $(z_3,z_4,z_5,z_6)$ along which $\mathbf{A}$ acts non-trivially), while the identity term has $\bvec[g]=0$ and hence satisfies \eqref{blorp2}. Therefore, although \eq{az4} does not descend to a well-defined diffeomorphism on $\mathcal{F}_7$, it acts in the Casimir and flux potentials as if it did. In particular, its action on the metric 
\begin{align}
    \mathbf{G}\,&\rightarrow \mathbf{A}^T\cdot\mathbf{G}\cdot\mathbf{A} \nonumber \\
    &=\left(
    \begin{array}{ccccccc}
         e^{u-2 w} & 0 & 0 & 0 & 0 & 0 & 0 \\
         0 & e^{u-2 w} & 0 & 0 & 0 & 0 & 0 \\
         0 & 0 & e^{u+w} \cosh (\zeta) & -\frac{e^{u+w} \sinh (\zeta)}{\sqrt{2}} & 0 & \frac{e^{u+w} \sinh (\zeta)}{\sqrt{2}} & 0 \\
         0 & 0 & -\frac{e^{u+w} \sinh (\zeta)}{\sqrt{2}} & e^{u+w} \cosh (\zeta) & -\frac{e^{u+w} \sinh (\zeta)}{\sqrt{2}} & 0 & 0 \\
         0 & 0 & 0 & -\frac{e^{u+w} \sinh (\zeta)}{\sqrt{2}} & e^{u+w} \cosh (\zeta) & -\frac{e^{u+w} \sinh (\zeta)}{\sqrt{2}} & 0 \\
         0 & 0 & \frac{e^{u+w} \sinh (\zeta)}{\sqrt{2}} & 0 & -\frac{e^{u+w} \sinh (\zeta)}{\sqrt{2}} & e^{u+w} \cosh (\zeta) & 0 \\
         0 & 0 & 0 & 0 & 0 & 0 & e^{-6 u} \\
    \end{array}
\right)
\end{align}
maps $\zeta\,\rightarrow-\zeta$, leaving $u$ and $w$ invariant. Hence, at $\zeta=0$ it becomes a $\mathbb{Z}_2$ symmetry; the component of the gradient $\partial_\zeta V$ transforms as a vector under this $\mathbb{Z}_2$, so that ensuring that the $G_4$ flux choice respects this symmetry will guarantee $\partial_\zeta V=0$ as desired. In practice, we will be more general and search for solutions with $G_4$ flux violating this symmetry; it turns out that the solution we find respects it.

To compute Casimir energies we must specify the spin structure on $\mathcal{F}_7$. The choices of spin structure are labelled by a vector $\vec{h}$ whose components are all 0 or 1/2 modulo 1, and encode the periodicity of fermions along the seven cycles of $T^7$ \cite{ValeixoBento:2025yhz}. The spin structures on $T^7$ that descend to well-defined spin structures on $\mathcal{F}_7$ are determined by solving the equations \eqref{eq:spin-structure-condition-1} and \eqref{eq:spin-structure-condition-2}.
Moreover, the spin lift $\mathcal{D}_\mathbf{g}$ of the generator satisfies $\mathcal{D}_\mathbf{g}^8 = \mathbf{I}$, i.e. $\mathfrak{s}_{\bf g} = 0 \,\text{mod}\,2\Z$, leading to the constraint $h_7 = 0$. For $\mathcal{F}_7$ there are then 4 possibilities for $\vec{h}$ compatible with the isometries discussed above, 
\begin{equation} 
    \left(0,0,0,0,0,0,0\right) \,,
    \left(\tfrac12,\tfrac12,0,0,0,0,0\right) \,,
    \left(0,0,\tfrac12,\tfrac12,\tfrac12,\tfrac12,0\right) 
    \,\,\text{or}\,\, 
    \left(\tfrac12,\tfrac12,\tfrac12,\tfrac12,\tfrac12,\tfrac12,0\right) \,.
\end{equation}
We will choose $\left(\tfrac12,\tfrac12,\tfrac12,\tfrac12,\tfrac12,\tfrac12,0\right)$, for which fermions are antiperiodic along all coordinates of the fibre, which will yield a dS saddle; we have also checked that the alternatives do not lead to better saddles.
Most cyclic RFMs do not allow for an antiperiodic spin structure on the subspace where the point group acts transitively; the only 7d cyclic RFMs that allow for antiperiodic boundary conditions, which are crucial to obtain a saddle point, have the same $\mathbb{Z}_8$ block as in \eq{Z8-rfmdef} \cite{ValeixoBento:2025yhz}. 

There is also some freedom in choosing the shift vector $\bvec[g]$. The allowed shifts are in one-to-one correspondence with solutions to the equation
\begin{equation}
    \D[g]\,\bvec[g]=\bvec[g]\quad\text{mod}\,\mathbb{Z} \,,
\end{equation}
which in our case has 4 inequivalent solutions,
\begin{equation}
    \left(0,0,0,0,0,0,0,\tfrac18\right) ,
    \left(\tfrac12,\tfrac12,0,0,0,0,\tfrac18\right) ,
    \left(0,0,\tfrac12,\tfrac12,\tfrac12,\tfrac12,\tfrac18\right) 
    \,\text{or}\, 
    \left(\tfrac12,\tfrac12,\tfrac12,\tfrac12,\tfrac12,\tfrac12,\tfrac18\right) \,,
    \label{eq:Z8-shift-vector}
\end{equation}
so that \eq{Z8-rfmdef} provides only a particularly simple choice of $\bvec[g]$. In fact, this is also the optimal choice, which we can conclude as follows. The shift vector's contribution to the solution comes from the enhancement of the Casimir potential computed on $\mathcal{F}_7$; therefore, the only effect of choosing a different $\bvec[g]$ will show up in the size of the Casimir contribution of each element of $\Z_8$. On the one hand, this means that the identity contribution is independent of $\bvec[g]$; on the other, only components of $\bvec[g]$ along the subspace invariant under $\D$ (i.e. each element $\gamma\in\Z_8$) will affect the corresponding twisted term in the Casimir potential. Since the directions $(z_3,z_4,z_5,z_6)$ never belong to the invariant subspace (recall that this block in $\D[g]$ has order $8$), these components of $\bvec[g]$ cannot affect the final solution. Moreover, the components along $(z_1,z_2)$ become relevant only for the element $\D[g]^4$ (recall that this block in $\D[g]$ has order $4$), but the contribution of this element to the Casimir potential vanishes when taking into account the traces over the M-theory field representations. As a consequence, the alternative choices of shift vector do not affect the stabilisation of the modulus $R$.
Hence, the particular choice \eq{Z8-rfmdef} is not only simple, but also equivalent to any of the others.

\subsection{Casimir energy}
\label{subsec:dS5-maximum-casimir-potential}

In four dimensions, the Casimir energy is a moduli-dependent quantity with units of length$^{-4}$ which is obtained from a one-loop calculation of the massless fields. Since the only dimensionful modulus is $R$, dimensional analysis forces a Casimir term of the form
\begin{equation}
    \Vcas = -\frac{\mathcal{C}(u,w,\zeta)}{R^4} \,,
    \label{casdef}
\end{equation}
where all non-trivial dependence is packed in a Casimir function $\mathcal{C}(u,w,\zeta)$ and with the overall minus sign taking into account the fact that Casimir energies of periodic fields are usually negative; hence, we expect $\mathcal{C}(u,w,\zeta)$ to be a positive function. 

\begin{figure}[!htb]
\centering
\begin{subfigure}{0.48\textwidth}
    \includegraphics[width=\textwidth]{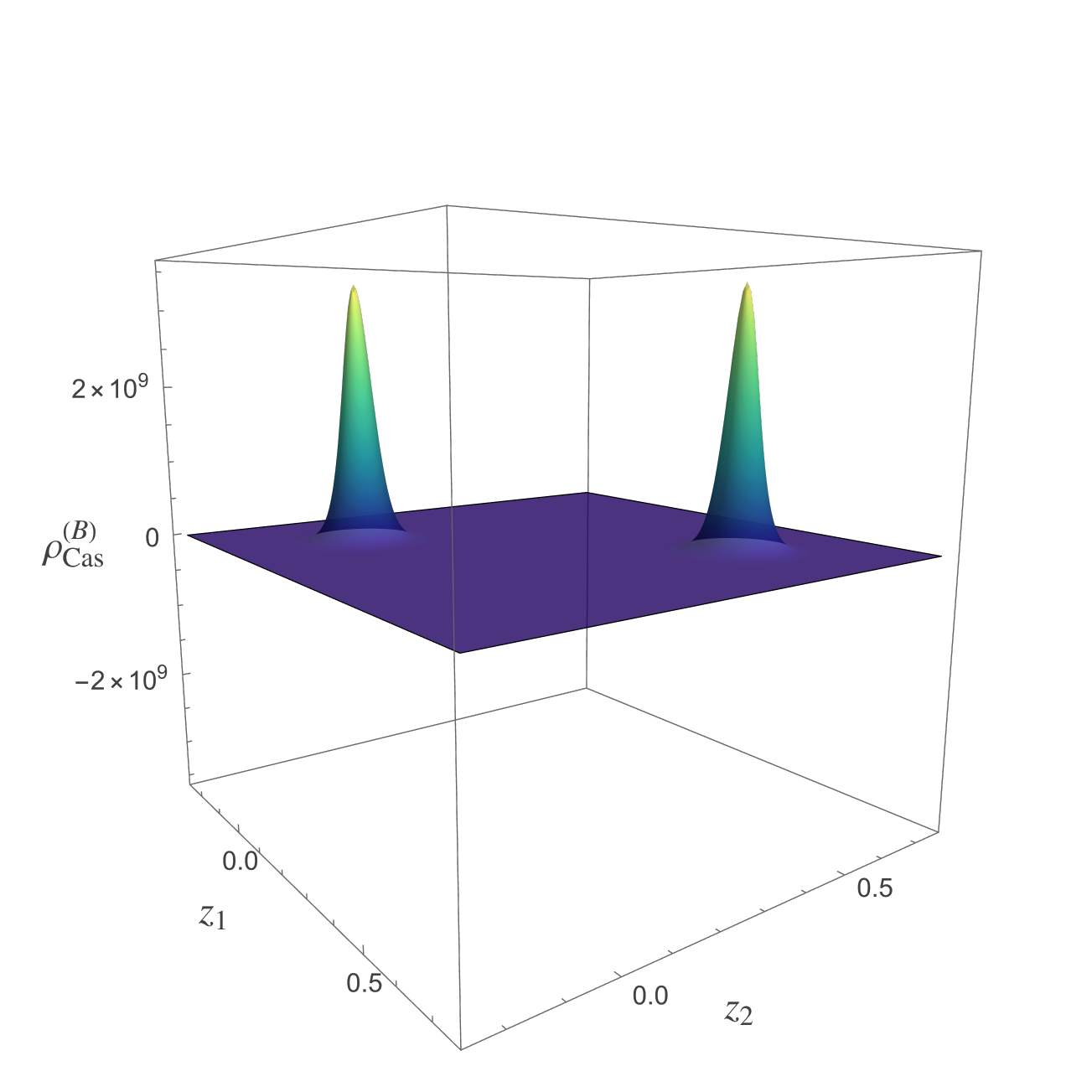}
\end{subfigure}
\hfill
\begin{subfigure}{0.48\textwidth}
    \includegraphics[width=\textwidth]{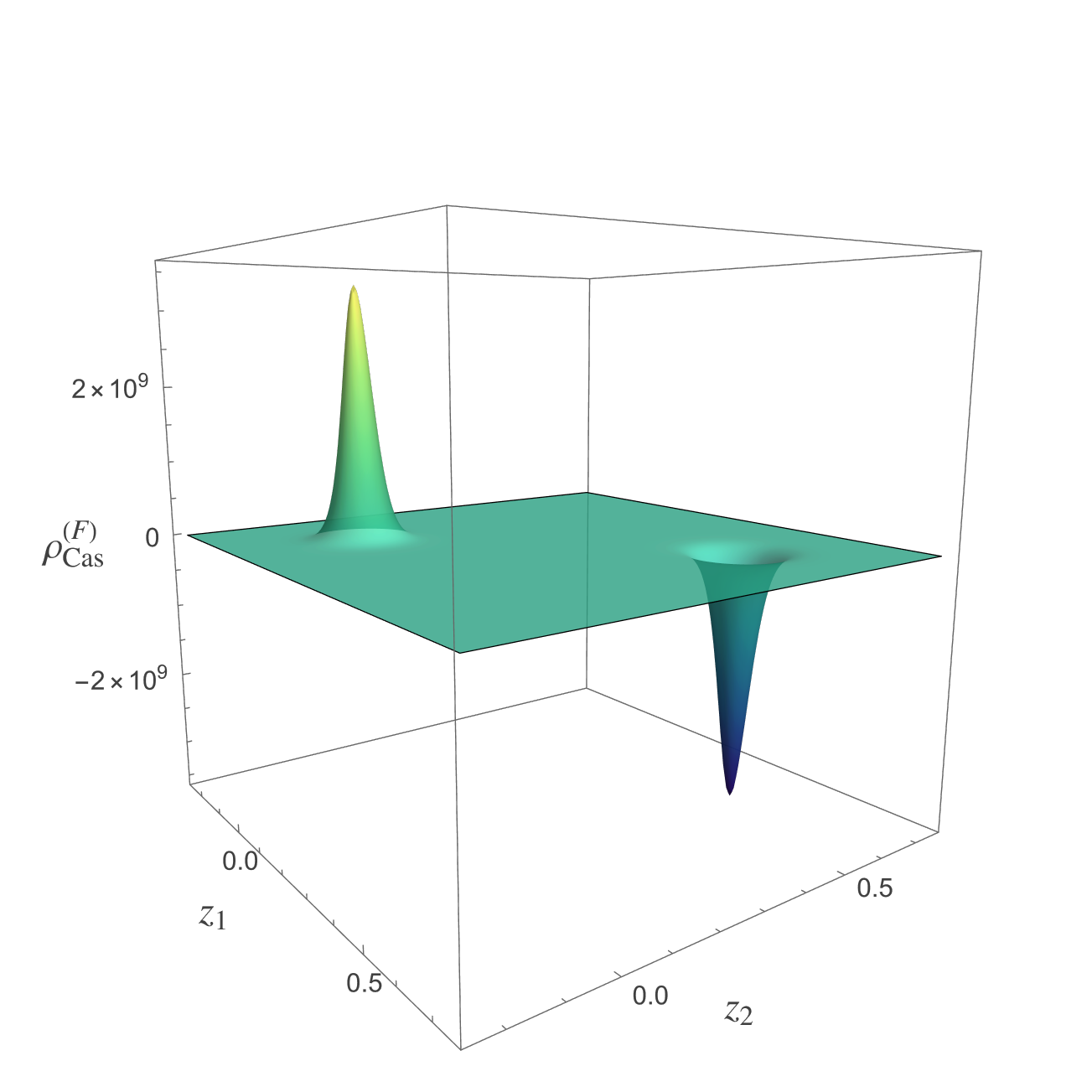}
\end{subfigure}
\begin{subfigure}{0.49\textwidth}
    \includegraphics[width=\textwidth]{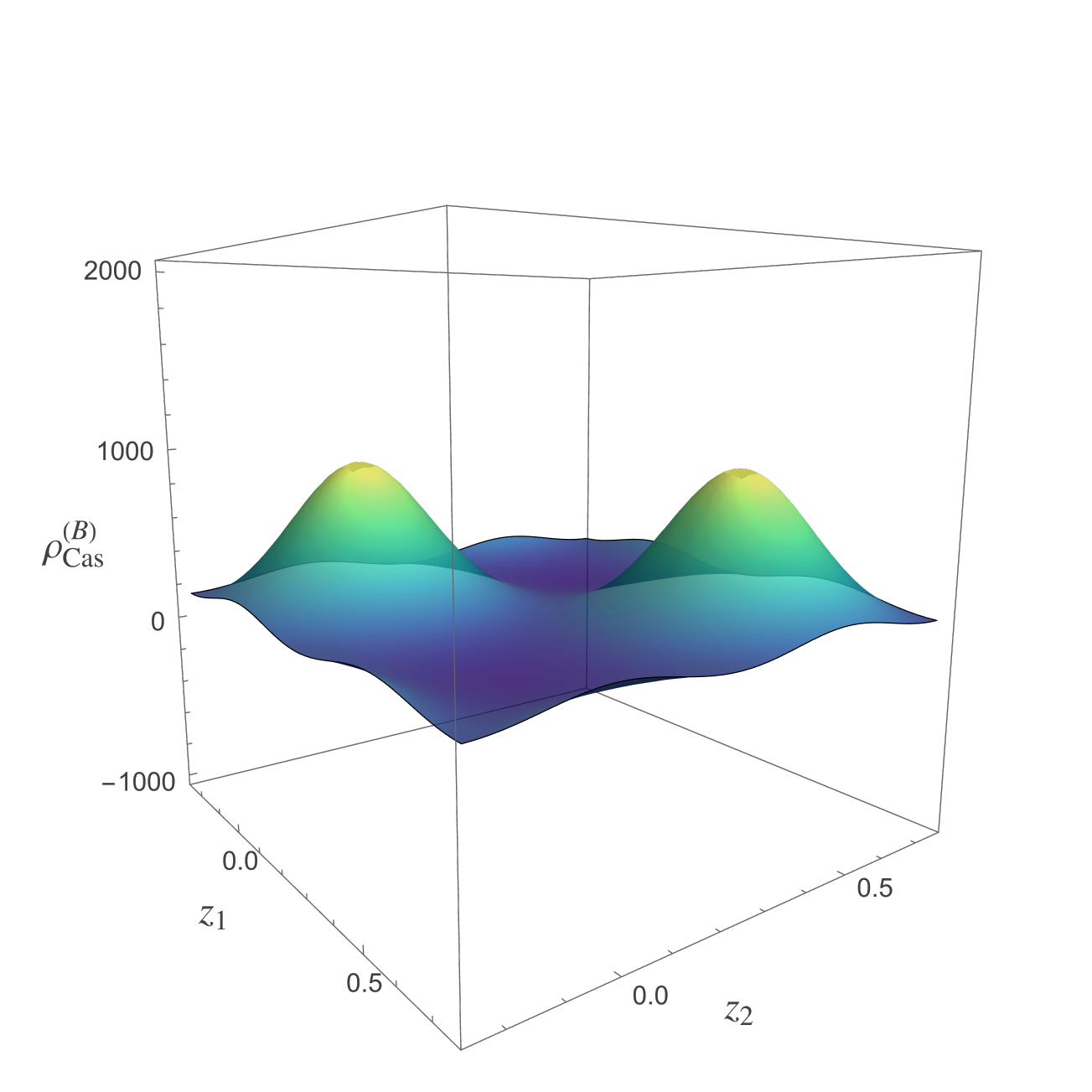}
\end{subfigure}
\hfill
\begin{subfigure}{0.49\textwidth}
    \includegraphics[width=\textwidth]{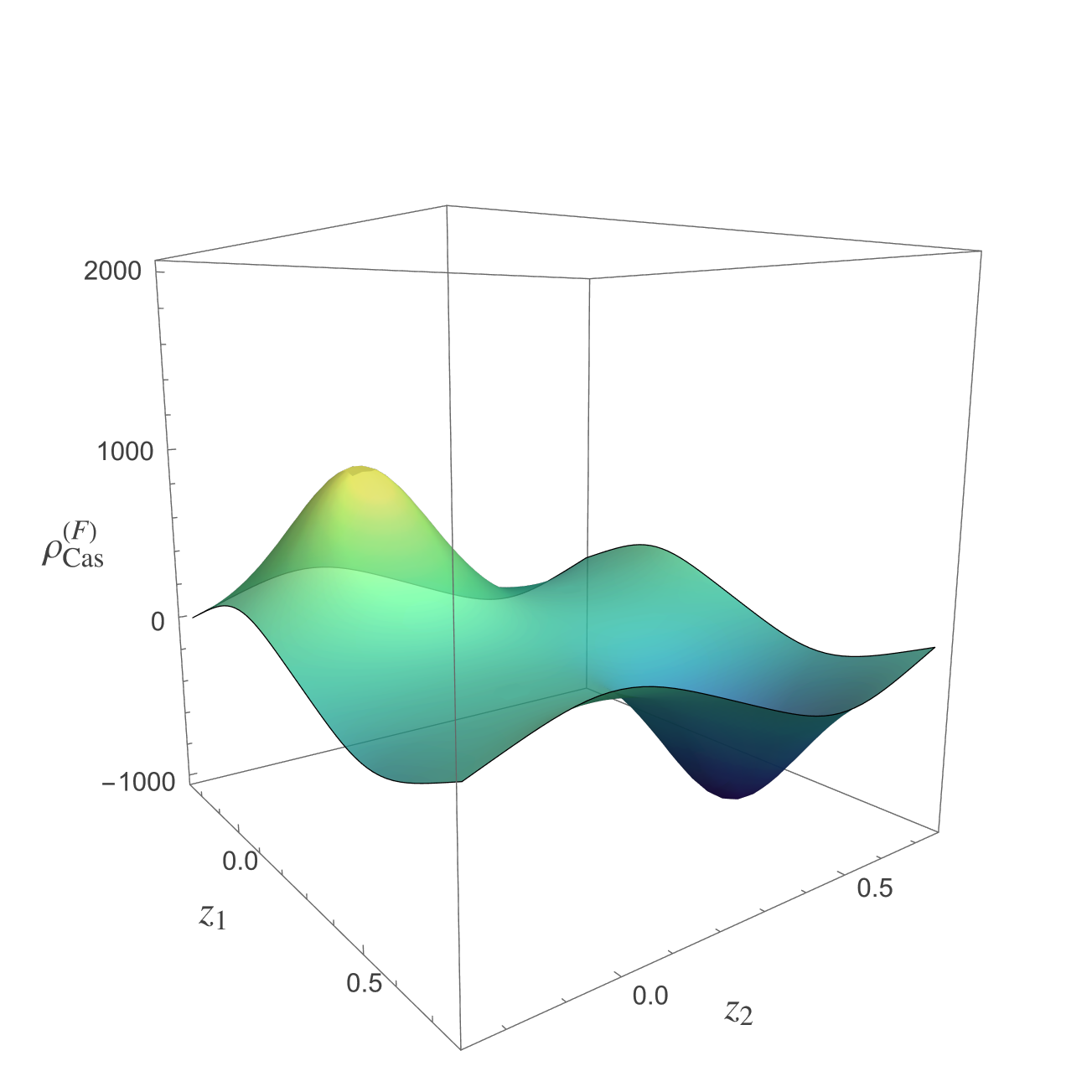}
\end{subfigure}
        
\caption{Contribution from the generator $\mathbf{g}\in\Z_8$ to the Casimir energy density of bosonic (left column) and fermionic (right column) degrees of freedom, with the fermionic boundary conditions being fixed by our choice of spin structure $\vec{h} = (\frac12,\frac12,\frac12,\frac12,\frac12,\frac12,0)$. By plotting over the $(z_1,z_2)$ plane we can see two Casimir branes, one at $(0,0,0,0,0,0)$ and one at $(\frac12,\frac12,0,0,0,0)$ on the fibre $T^6$ that wrap the base $S^1$ (there are also Casimir branes that we cannot see in this plot). While for bosons both Casimir branes have positive tension, for fermions they have opposite tension and cancel each other out upon integration over the covering $T^7$. We compare the magnitude of the Casimir branes at the ``square'' point in moduli space with $(u,w,\zeta)=(0,0,0)$ (top row) and at the point where we find the $dS_4$ saddle, $(u,w,\zeta)\approx (-0.457,0.0396,0)$ (bottom row); at the saddle point the Casimir energy density is smaller than at the ``square'' point by several orders of magnitude.}
\label{fig:Casimir-branes-Z8}
\end{figure}

The Casimir energy $\Vcas$ on $\mathcal{F}_7 = T^7/\Z_8$ can be computed from \eqref{eq:Casimir-general} by taking $\Gamma = \Z_8$, $\mathbf{r}\in\{\mathbf{44},\mathbf{84},\mathbf{128}\}$ (i.e. the graviton, 3-form and gravitino in M-theory) and $s=\frac{11}{2}$. The elements $\gamma\in\Z_8$ can be labelled with an index $j=0\,,\ldots,7$ such that $\mathbf{D}_j=\D[g]^j$ and $\vec{b}_j=j\,\bvec[g]$; each element contributes as a number of Casimir branes wrapping the subspaces invariant under $\mathbf{D}_j$, i.e. $\mathbf{D}_j\,\vec{z} = \vec{z}~\text{mod}~\Z^7$, and localised on orthogonal directions. While the identity term $(j=0)$ contributes as a space-filling brane wrapping the whole of $\mathcal{F}_7$, the order-2 element $(j=4)$ corresponds to 16 codimension-4 branes localised on the $T^4$ factor of the fibre and wrapping the remaining $T^2$ and the base $S^1$. All other elements contribute as a number of codimension-6 Casimir branes, localised on the $T^2\times T^4$ fibre and wrapping the base $S^1$. In total, the eight elements of $\Z_8$ contribute as 65 Casimir branes of various codimensions.  
A crucial property of Casimir energies computed on RFMs is encoded by the factor $\hat{\delta}_{\vec{h}}$ in \eqref{eq:Casimir-general}, determined by the equation \eqref{eq:eta-condition}. This factor was interpreted in \cite{ValeixoBento:2025yhz} as a spacetime Atkin-Lehner-like symmetry \cite{atkin1970hecke,Moore:1987ue,Dienes:1990qh} and arises due to a cancellation between positive and negative tension Casimir branes upon integration over the RFM (see Fig.~\ref{fig:Casimir-branes-Z8}). For $\mathcal{F}_7$, we have $\hat{\delta}_{\vec{h}}=0$ for all fermionic contributions of non-trivial (twisted) elements, meaning that the gravitino only contributes to the Casimir potential through the identity element $(j=0)$.

\begin{table}[t]
    \centering
    \renewcommand{\arraystretch}{1.2}
    \begin{tabular}{c|cccccccc}
         $j$ & 0 & 1 & 2 & 3 & 4 & 5 & 6 & 7  \\ \hline 
        $\Tr{\mathbf{44}}{\D[g]^j}$ & 44 & 4 & 0 & 4 & 4 & 4 & 0 & 4  \\
        $\Tr{\mathbf{84}}{\D[g]^j}$ & 84 & 4 & 0 & 4 & $-4$ & 4 & 0 & 4  \\
        $\Tr{\mathbf{128}}{\D[g]^j}$ & 128 & $\pm 8$ & 0 & $\pm 8$ & 0 & $\pm 8$ & 0 & $\pm8$ 
    \end{tabular}
    \caption{Traces of each element $\D[g]^j\in\Z_8$ in the graviton, 3-form and gravitino representations. The signs of traces in the gravitino representation are fixed by the choice of spin lift of the generator of $\Gamma$; our choice of lift below is such that these traces are negative.}
    \label{tab:traces-Z8}
\end{table}
The traces of each element in the graviton, 3-form and gravitino representations can be obtained using the formulas in Appendix F of \cite{ValeixoBento:2025yhz} (see Table \ref{tab:traces-Z8}). Note that the traces over bosonic representations vanish for all even non-trivial elements, so that these do not contribute at all to the Casimir potential; this includes in particular the contribution from the $j=4$ element, corresponding to codimension-4 Casimir branes. 
The full twisted contribution to the Casimir potential is then
\begin{align}
    \mathcal{C}_\text{twisted}(u) &= \frac{\Gamma(\tfrac52)}{2\pi^{5/2}}\cdot 2\,e^{12\,u}\bigg\{
        \sum_{k\in\Z} \frac{1}{|k - \frac{1}{8}|^{5}}
        + \sum_{k\in\Z} \frac{1}{|k - \frac{3}{8}|^{5}}
    \bigg\} \nonumber \\
    &= \frac{\Gamma(\tfrac52)}{2\pi^{5/2}}\cdot 2\,e^{12\,u}\Big\{
        \zeta_{\rm H}\big(5,\tfrac18\big) + \zeta_{\rm H}\big(5,\tfrac38\big) + \zeta_{\rm H}\big(5,\tfrac58\big) + \zeta_{\rm H}\big(5,\tfrac78\big)
    \Big\}\,,
\end{align}
where $\zeta_{\rm H}(a,b)$ is the Hurwitz zeta function,
arising from the bosonic contribution of non-trivial odd elements with $j\in\{1,3,5,7\}$, which collapse into one-dimensional sums over the base $S^1$ corresponding to the invariant sublattice under these $\D[g]^j$. Note that we have used the metric \eqref{eq:RFM-metric}, as well as $s_j = \frac52$ and $\Tr{44}{\mathbf{D}_j} + \Tr{84}{\mathbf{D}_j} = 8$ for $j=1,3,5,7$; we have also exploited the symmetry of the sums over $k$ under $j\to 8-j$ in order to pair these elements in the first line. 
The twisted terms only generate a Casimir potential for the modulus $u$ (in addition to the overall power of $R$ in \eqref{casdef}), with a particularly simple exponential dependence. 

Finally, since the full $\Z^7$ lattice remains invariant under the identity element $\D[g]^0 = \mathbf{I}$, its contribution to the Casimir energy remains a full 7d infinite lattice sum that must be computed numerically,
\begin{equation}
    \mathcal{C}_\mathbf{I}(u,w,\zeta) = \frac{\Gamma(\tfrac{11}{2})}{16\pi^{11/2}}\cdot 128
    \sum_{\substack{(\vec{n},\vec{m},k)\in\Z^7}} \frac{1 - e^{\pi i (n_1+n_2+m_1+m_2+m_3+m_4)}}{\Big[e^{u-2w}\,|\vec{n}|^2 + e^{u+w}\,|\vec{m}|_\zeta^2 + e^{-6u}\,k^2 \Big]^{\tfrac{11}{2}}} \,,
    \label{eq:Casimir-dS4-Id}
\end{equation}
and depends on all shape moduli. It can be computed using the methods introduced in \cite{ValeixoBento:2025yhz} and implemented in \cite{ValeixoBento:2026jdx}.
We can now see explicitly that $\zeta\to-\zeta$ is indeed a symmetry of the Casimir potential, since
\begin{equation}
    |\vec{m}|^2_\zeta = 
    \cosh(\zeta)\big[m_1^2 + m_2^2 + m_3^2 + m_4^2\big] 
    + \sqrt{2}\sinh(\zeta)\big[m_1(m_2-m_4) + m_3(m_2+m_4)\big] \,,
\end{equation}
and the $\Z_2$ action can be compensated through sign flips $m_1\to -m_1$ and $m_3\to -m_3$ that do not change the sum over $\vec{m}\in\Z^4$. Therefore, $\zeta = 0$ must be a critical point of this potential.

Note that the asymptotic behaviour of $\mathcal{C}_\textbf{I}(u,w,\zeta)$ as $u\to\pm\infty\,,\,w\to\pm\infty$ depends on the spin structure. For example, as $u\to +\infty$, the leading terms in the sum \emph{would} correspond to $\vec{n}=\vec{m}=0$; however, this term vanishes for our choice of spin structure and the true asymptotic behaviour comes from subleading terms in the sum that give $\mathcal{C}_\textbf{I}\sim e^{3\,u}$ as $u\to+\infty$. Conversely, in the opposite limit $u\to -\infty$, the leading terms in the sum correspond to $\vec{m}=0$ and $k=0$, which no longer cancels and leads to the asymptotic behaviour $\mathcal{C}_\textbf{I}\sim e^{-\frac{11}{2}\,u}$ as $u\to-\infty$. A similar analysis applies to the $w$ asymptotics: the leading terms in the sum as $w\to+\infty$ have $\vec{m}=0$ and $k=0$, and we find the asymptotic behaviour $\mathcal{C}_\textbf{I}\sim e^{11\,w}$; the leading terms in the opposite limit $w\to-\infty$ have $\vec{n}=0$ and $k=0$, with an asymptotic behaviour $\mathcal{C}_\textbf{I}\sim e^{-\frac{11}{2}\,w}$. 

\subsection{Flux potential}
\label{subsec:dS4-maximum-flux-potential}

In our setup, the flux piece $V_{G_4}$ of the scalar potential \eqref{totpot} comes entirely from M-theory $G_4$ flux. The $G_4$ kinetic term is proportional to the Hodge norm $\vert G_4\vert^2$, which we now compute. To begin with, we must take into account that $G_4$ is quantised,
\begin{equation}
    \int_{\mathcal{C}_4}G_4\in\mathbb{Z}\,,\quad\forall\, \mathcal{C}_4\in H_4(\mathcal{F}_7,\mathbb{Z}) \,,
    \label{uf}
\end{equation}
which can be equivalently written as 
\begin{equation}
    \int_{\mathcal{F}_7}\omega\wedge G_4\in\mathbb{Z}\,,\quad\forall\, \omega\in H^3(\mathcal{F}_7,\mathbb{Z}) \,,
    \label{uf2}
\end{equation}
using Poincar\'{e} duality. Choosing a basis $\{\omega_i\}$ of $H^3(\mathcal{F}_7,\mathbb{Z})$, we may satisfy \eq{uf} and \eq{uf2} by expanding $G_4$ in terms of the Hodge-dual forms $\{*\omega_i\}$---that do not, in general, have quantised periods---as
\begin{equation} 
    G_4=\sum_{i,j} n^i\, \mathcal{G}_{ij}^{-1} *\omega_j\,,\quad n^i\in\mathbb{Z} \,,
\end{equation}
where we have defined 
\begin{equation}
    \mathcal{G}_{ij}\equiv \int_{\mathcal{F}_7} \omega_i\wedge *\omega_j \,.
    \label{kinterm}
\end{equation}
With these definitions, we have 
\begin{equation} 
    \int_{\mathcal{F}_7}\vert G_4\vert^2=\sum_{i,j} \mathcal{G}_{ij}^{-1}\, n^in^j\,.
    \label{fluxpoty}
\end{equation}
In order to determine the Hodge norm acting on 3-forms \eq{kinterm}, we must construct a basis of $H^3(\mathcal{F}_7,\mathbb{Z})$, and the natural projection map $\pi:\, T^7\rightarrow\mathcal{F}_7$ allows us to work in the covering $T^7$. Explicitly, we have
\begin{equation} 
    \int_{\mathcal{F}_7} \omega_i\wedge *\omega_j=  \frac18\int_{T^7} \pi^*(\omega_i)\wedge *\pi^*(\omega_j)\,,
    \label{kinterm2}
\end{equation}
where the factor of $1/8$ appears because the fundamental class of $\mathcal{F}_7$ pulls back to 8 times the fundamental class of $T^7$ (in other words, $T^7$ is an eightfold covering of $\mathcal{F}_7$, which can be identified with a subset of $T^7$ with 1/8 its volume). The pullback $\pi^*(\omega_i)\in H^3(T^7,\mathbb{Z})$ is represented by a 3-form on $T^7$ invariant under the action \eq{Z8-rfmdef}. Out of the thirty-five 3-cycles of $T^7$, only a five-dimensional subspace is invariant under \eq{Z8-rfmdef}, and is generated by the 3-forms
\begin{align}
    \omega_1 &= 
    dz_3 \wedge dz_4 \wedge dz_7 
    + dz_3 \wedge dz_6 \wedge dz_7 
    + dz_4 \wedge dz_5 \wedge dz_7 
    + dz_5 \wedge dz_6 \wedge dz_7 \,, \nonumber \\
    \omega_2 &= 
    dz_3 \wedge dz_5 \wedge dz_7 
    + dz_4 \wedge dz_6 \wedge dz_7 \,, \nonumber \\
    \omega_3 &= 
    dz_1 \wedge dz_2 \wedge dz_7 \,, \nonumber \\
    \omega_4 &= 
    dz_1 \wedge dz_3 \wedge dz_6
    - dz_1 \wedge dz_4 \wedge dz_5
    + dz_2 \wedge dz_3 \wedge dz_4
    - dz_2 \wedge dz_5 \wedge dz_6 \,, \nonumber  \\
    \omega_5 &= 
    -\,dz_1 \wedge dz_3 \wedge dz_4
    + dz_1 \wedge dz_5 \wedge dz_6
    + dz_2 \wedge dz_3 \wedge dz_6
    - dz_2 \wedge dz_4 \wedge dz_5 \,.
    \label{invariant2forms}  
\end{align}
Note that $\omega_1\to -\omega_1$, $\omega_4\to -\omega_4$ and $\omega_5\to -\omega_5$ under \eqref{az4}, while $\omega_2$ and $\omega_3$ are invariant. 
The sublattice $\pi^*H^3(\mathcal{F}_7,\Z)$ with properly quantised periods is generated by $\{2\,\omega_1,4\,\omega_2,8\,\omega_3,\omega_4,\omega_5\}$, as shown in Appendix \ref{app:A}.

The metric $\mathcal{G}_{ij}$ then becomes 
\begin{equation}
    \mathcal{G}_{ij}^{-1} = \left(
        \begin{array}{ccccc}
         \frac{1}{2}\,e^{2 w-4 u} \cosh (2 \zeta) & \frac{1}{2\sqrt{2}}\,e^{2 w-4 u}
           \sinh(2\zeta) & 0 & 0 & 0 \\
         \frac{1}{2\sqrt{2}}\,e^{2 w-4 u} \sinh(2\zeta) &
           \frac{1}{4}\,e^{2 w-4 u} \cosh (2 \zeta) & 0 & 0 & 0 \\
         0 & 0 & \frac{1}{8}\,e^{-4(u+w)} & 0 & 0 \\
         0 & 0 & 0 & 2 e^{3 u} & 0 \\
         0 & 0 & 0 & 0 & 2 e^{3 u} \\
        \end{array}
    \right) \,,
    \label{eq:flux-metric}
\end{equation}
which can be computed using \eq{kinterm2} together with the properly quantised basis $\pi^*H^3(\mathcal{F}_7,\Z)$ determined above.
From \eq{fluxpoty}, we finally obtain our flux potential,
\begin{align}\label{eq:VG4}
    V_{G_4}
    &= \frac{(2\pi^2)^{\frac{1}{3}}}{2\ell_{11}^3\,R}\sum_{i,j} \mathcal{G}^{-1}_{ij}n_in_j \\
    &= \frac{(2\pi^2)^{\frac{1}{3}}}{2\ell_{11}^3\,R} \bigg(
        \,e^{2 w-4 u}\Big(\frac{2n_1^2+n_2^2}{4} \cosh(2\zeta)
        +\frac{n_1n_2}{\sqrt{2}}\sinh(2\zeta)\Big)
        + \frac{n_3^2}{8}\,e^{-4(u+w)} 
        + 2e^{3u}(n_4^2 + n_5^2)
    \bigg) \, .\nonumber 
\end{align}
We can immediately see that if $n_1n_2 = 0$ (i.e. if these fluxes are not simultaneously turned on), the flux potential respects the symmetry \eqref{az4} and thus the critical point at $\zeta = 0$.
Furthermore, note that the $\Z_8$ quotient of the RFM reduces the physical flux increment associated with a unit quantum (in particular for $n_1,n_2,n_3$ fluxes) allowing for finer choices of flux that lead to more and better vacua in the parent torus. 
%Due to the structure of the two-term potential, better solutions (i.e. with larger volumes and smaller vacuum energy, and thus in better control) are expected to arise from the smallest flux contribution. From these considerations, one might expect $n_3 = 1$ to be the most natural candidate. However, such a choice does not lead to a saddle. Instead, 
Indeed, in the next section we present a $dS_4$ solution on the RFM given by \eqref{Z8-rfmdef} arising for $n_2=1$ and all other fluxes turned off.

\subsection{The \texorpdfstring{$dS_4$}{dS4} maximum}
\label{subsec:dS4-maximum}

The full scalar potential $V$ is simply obtained by combining the results of Subsections \ref{subsec:dS5-maximum-casimir-potential} and \ref{subsec:dS4-maximum-flux-potential}. Adding the flux \eqref{eq:VG4} and Casimir \eqref{casdef} contributions, we find 
\begin{align}
    \frac{V^{(4d)}}{\MPl^4} = \left(\frac{8}{R^7}\right)^{2} \Bigg[&
    \frac{(2\pi^2)^{\frac{1}{3}}}{2\,R} \bigg\{
    \frac{2n_1^2+n_2^2}{4}\,e^{2 w-4 u} \cosh (2 \zeta)
    + \frac{n_3^2}{8}\,e^{-4(u+w)}+ 2e^{3u}(n_4^2 + n_5^2) \nonumber \\
    &\hspace{5em}+\frac{n_1n_2}{\sqrt{2}}\,e^{2 w-4 u}\sinh(2\zeta)
    \bigg\} 
    - \frac{\mathcal{C}(u,w,\zeta) 
    }{R^4} \Bigg] \,,
    \label{eq:full-potential}
\end{align}
where we have also included an overall Weyl rescaling factor, required to go to Einstein frame in 4d, and set $\ell_{11} = 1$. Note that, while the potential is in 4d Planck units, the dimensionful modulus $R$ is in units of $\ell_{11}$. We can now look for extrema of this potential---for the most general choice of fluxes, both terms depend on all moduli, so it is not very useful to write analytical expressions in general. On the other hand, for potentials of the form \eqref{eq:full-potential}, we can always solve for $R$ analytically and reduce the problem to the extremisation of a potential that only depends on the remaining moduli. In this case, we can define
\begin{align}
    V_R (u,w,\zeta)\equiv \frac{5^5}{3^6\,\mathcal{C}(u,w,\zeta)^5}\Bigg[
    \frac{(2\pi^2)^{\frac{1}{3}}}{2} \bigg\{&
    \frac{2n_1^2+n_2^2}{4}\,e^{2 w-4 u} \cosh (2 \zeta)
    + \frac{n_3^2}{8}\,e^{-4(u+w)}+ 2e^{3u}(n_4^2 + n_5^2) \nonumber \\
    &\hspace{5em}+\frac{n_1n_2}{\sqrt{2}}\,e^{2 w-4 u}\sinh(2\zeta)
    \bigg\} \Bigg]^6 \,,
\end{align}
after using the solution for $R$ given by 
\begin{align}
    R^3 = \frac{6\,\mathcal{C}(u,w,\zeta)}{5}\Bigg[
    \frac{(2\pi^2)^{\frac{1}{3}}}{2} \bigg\{&
    \frac{2n_1^2+n_2^2}{4}\,e^{2 w-4 u} \cosh (2 \zeta)
    + \frac{n_3^2}{8}\,e^{-4(u+w)}+ 2e^{3u}(n_4^2 + n_5^2) \nonumber \\
    &\hspace{5em}+\frac{n_1n_2}{\sqrt{2}}\,e^{2 w-4 u}\sinh(2\zeta)
    \bigg\} \Bigg]^{-1} \,.
    \label{eq:R-solution}
\end{align}
We want to find the solution with the largest possible $R$, since this will give us the most control. The solution for $R$ \eq{eq:R-solution} favours the largest possible $\mathcal{C}(u,w,\zeta)$ and the smallest allowed fluxes, both of which also make the vacuum energy the smallest; however, the influence of the fluxes on the vevs of the remaining moduli is less straightforward and could invalidate this intuition.
For $n_2 = 1$, with all other fluxes turned off, we find a solution with $R\sim 5.16$ and vacuum energy $V_\text{saddle} = 5.02\cdot 10^{-10}\,\MPl^4$ (see Table \ref{tab:dS4-Z8-solution}). Although we have not performed a fully systematic scan, we have checked that other solutions are either worse or not significantly different from this one.

\begin{table}
    \centering
    \renewcommand{\arraystretch}{1.5}
    \begin{tabular}{ccccccccc}
        \hline
        $R$ & $u$ & $w$ & $\zeta$ & $R\,e^{\frac{u}{2}-w}$ & $R\,e^{\frac12(u+w)}$ & $\frac18\,R\,e^{-3u}$ & $\mathrm{vol}(\mathcal{F}_7)$ & $V_{\rm saddle}$ \\ \hline
        5.16 & $-0.457$ & 0.0396 & 0 & 3.94 & 4.18 & 2.54 & $1.21\cdot 10^4$ & $5.02\cdot 10^{-10}$ 
        \\ \hline
    \end{tabular}
    \caption{Parameters for the saddle point solution found for $(n_1,n_2,n_3,n_4,n_5) = (0,1,0,0,0)$. We also list the length of the smallest curves in the base and fibre of $\mathcal{F}_7$, in 11-dimensional Planck units, as well as the volume of the space; the vacuum energy is given in 4-dimensional Planck units.}
    \label{tab:dS4-Z8-solution}
\end{table}

Although this is indeed a small vacuum energy, it still falls short of the hope that large RFM Casimir energies would lead to extremely well-controlled solutions. There are a few reasons for this---first of all, just like in the $dS_5$ solution of \cite{ValeixoBento:2025yhz}, this $dS_4$ saddle arises from balancing twisted and untwisted terms in the sum \eqref{eq:Casimir-general}; secondly, even the twisted terms are not contributing with their expected large numbers. A na\"ive estimate of the twisted term contribution would give 
\begin{equation}
    \mathcal{C}_{\rm twisted}\sim \frac{1}{|\Gamma|}\sum_{\gamma\in\Gamma}\,\min_{\vec{n}\in\Z^7}|\bvec+\vec{n}|^{-4-D_\text{inv}(\gamma)} \sim 8.5 \cdot 10^{3} \,,
    \label{eq:Ctwisted-estimate}
\end{equation}
where $D_\text{inv}$ is the dimension of the invariant subspace under $\D$ and $\bvec$ the shift vector of the element $\gamma\in\Gamma$. In our present case, $\Gamma = \Z_8$ and our expectation for twisted terms is of order $10^4$. However, this estimate disregards the effect of moduli values, in essence assuming a ``square'' RFM---in practice, the moduli acquire vevs that can be relatively far from this ``square'' regime and affect the estimate quite significantly (see Fig.~\ref{fig:Casimir-branes-Z8}). In our case, this is directly tied to the expectation value of the modulus $u$, which appears as $e^{-6u}$ multiplying the term with the shift vector in the Casimir sum; rather than $\bvec$ itself, what ends up contributing to the sum at our saddle point is $e^{-3u}\bvec\sim 4\bvec$. Using this instead in our estimate \eqref{eq:Ctwisted-estimate}, we would find $\mathcal{C}_\text{twisted}\sim 16$, roughly 3 orders of magnitude smaller. In fact, since our Casimir coefficient turns out to be $\mathcal{C}\sim 260$, we should expect most of this to come from the identity term; this is indeed what we find (see Table \ref{tab:Casimir-by-element}). 

\begin{table}[]
    \centering
    \renewcommand{\arraystretch}{1.2}
    \begin{tabular}{c|cccccccc|c}
         $j$ & 0 & 1 & 2 & 3 & 4 & 5 & 6 & 7 & Total \\ \hline 
        $\mathcal{C}(-0.457,0.0396,0)$ & 249 & 5.17 & 0 & 0.023 & 0 & 0.023 & 0 & 5.17 & 260  \\
        $\mathcal{C}(0,0,0)$ & 20.5 & 1245 & 0 & 5.53 & 0 & 5.53 & 0 & 1245 & 2522
    \end{tabular}
    \caption{Contribution of each element $\D[g]^j\in\Z_8$ to the Casimir potential, both at our saddle point and at a reference ``square'' point with $(u,w,\zeta)=(0,0,0)$.}
    \label{tab:Casimir-by-element}
\end{table}

We have also included in Table \ref{tab:dS4-Z8-solution} the lengths of the shortest representative one-cycles of $\mathcal{F}_7$, which will either be entirely contained in the $T^2$ block, the $T^4$ block or in the $S^1$ base. We can read off the sizes of physical closed curves on $\mathcal{F}_7$ from the metric \eqref{eq:RFM-metric},
\begin{equation} 
    R\,e^{\frac{u}{2}-w} \,,\quad
    R\,e^{\frac12(u+w)} \,,\quad\text{and}\quad 
    \frac18\,R\,e^{-3u} \,,
\end{equation}
where we have included the factor of $\frac18$ on the cycle along the base due to the $\Z_8$ quotient. These are the parameters quoted in Table \ref{tab:dS4-Z8-solution}.
Although the value of the vacuum energy is quite small, the size of the shortest cycles is still Planckian, which raises issues of control that we systematically discuss in what follows. 

\subsection{Characterisation and control of the solution}
\label{sec:ctrl}

In the previous sections we have found a saddle point solution of M-theory on $dS_4\times\mathcal{F}_7$, in the form of a Riemann-flat compactification on $\mathcal{F}_7=T^7/\Z_8$, arising from the balance of $G_4$ flux and a Casimir energy contribution. Our solution has $n_2=1$ unit of $G_4$ flux along the fibre, and all other components turned off, and results in a small vacuum energy of $5.02\cdot 10^{-10}$ in four-dimensional Planck units. Let us now provide a more detailed description of this solution, and carefully analyse the amount of control it provides.

The small vacuum energy leads to a dS radius, or Hubble scale,
\begin{equation}
    H_0^{-1}=\sqrt{\frac{3}{V_{\text{saddle}}}}\approx 7.73\cdot 10^4 \,\ell_4 \,.
    \label{hradius}
\end{equation}
The ratio of the 4d and 11d Planck lengths is
\begin{equation}
    \frac{\ell_{11}}{\ell_4} = \sqrt{\frac{R^7}{8}}\approx 1.10\cdot 10^2 \,, 
\end{equation}
so the solution is scale separated and the four-dimensional EFT is valid up to roughly $10^{-2}\, \MPl$.

Determining the spectrum of light fields and masses around this solution requires the properly normalised kinetic terms of the moduli in the RFM. Since RFMs are quotients of tori, the kinetic terms from dimensional reduction of the $(D=d+k)$--dimensional theory on $T^k$ are enough. The kinetic terms of the moduli coming from the Einstein-Hilbert action on $T^k$ are
\cite{Etheredge:2022opl}
\begin{equation}
    \frac{1}{2\kappa_d^2}\int d^dx\, \sqrt{-g_d}\left[\frac{1}{(d-2)} (\partial\log\sqrt{\text{det}\,\mathbf{G}})^2+ \frac{1}{4}\text{Tr}((\mathbf{G}^{-1}\partial\mathbf{G})^2)\right] \,,
\end{equation}
and they can be used to properly normalise the Hessian matrix, whose eigenvalues give the squared masses of the moduli fields $\{R,u,w,\zeta\}$, 
\begin{equation}
    \{-35.3\,,\,-12\,,\,-323\,,\,109.8\}\cdot H_0^2 \,.
    \label{eq:dS4-masses}
\end{equation}
We conclude that our solution is a stable minimum along the $\zeta$ direction and an unstable maximum along the remaining three; the masses acquired by the four moduli are of the order of the Hubble scale $H_0\approx 1.29\cdot 10^{-5}\, \MPl$ \eqref{hradius}. 

On top of these geometric moduli, we get one KK vector from the base $S^1$, and additional fields from the dimensional reduction of $C_3$. The free part of the $\mathcal{F}_7$ cohomology is 
\begin{equation}
    \begin{array}{c|cccccccc}
        p&0&1&2&3&4&5&6&7\\\hline H^p(\mathcal{F}_7,\mathbb{R})&\mathbb{R}&\mathbb{R}&3\mathbb{R}&5\mathbb{R}&5\mathbb{R}&3\mathbb{R}&\mathbb{R}&\mathbb{R}
    \end{array}
\end{equation}
so from the dimensional reduction of $C_3$ we get three other vectors and six axions, five from $C_3$ periods and another one from the four-dimensional dual of $C_3$ with one leg on $T^7$ (on the base $S^1$). All of these fields are massless at leading order; the axions constitute compact directions of the moduli space, which receive a non-perturbative potential due to M2 and M5 brane instantons \cite{Harvey:1999as}. While we have not computed their exact minimum, the smallness of the potential together with the compact range of the scalars ensures that they give a negligible correction to the solution, and can be treated in practice as a moduli space. The shift symmetry of the axions ensures that there are no perturbative corrections \cite{Dine:1986vd}, with one exception: the $G_4$ flux leads to a $BF$ coupling in four dimensions and therefore one of the axions (the one coming from $C_3$ with one leg along $T^7$) is eaten by a vector, which becomes a massive vector. In some cases, one can use the parity symmetry of M-theory to ensure that the exact $C_3$-axion potential will have a minimum at zero; it is not possible to do this for all components of the axion fields in this case, due to the $G_4$ flux that breaks some of these parity symmetries.

Finally, there are no light fermions in our solution, since $\mathcal{F}_7$ is the quotient of a $T^7$ where we have chosen an antiperiodic spin structure around six of the seven 1-cycles. This means that the Dirac operator has no zero modes, nor does the Rarita-Schwinger field since the tangent bundle of $T^7$ is flat. As a result, our four-dimensional theory is purely bosonic, even though we started with a maximally supersymmetric theory in 11 dimensions.

Since we have a dS solution with massless vector fields, we could check the Festina Lente (FL) bound \cite{Montero:2019ekk}, which states that the masses of all charged states under any $U(1)$ gauge field must satisfy
\begin{equation}
    m^2\gtrsim \frac{g\,q}{\sqrt{G}} H \,,
    \label{fl}
\end{equation}
in terms of the Hubble scale $H$, the gauge coupling $g$, the integer quantised charge $q$ of the particle, and Newton's constant $G$. 
Strictly speaking, the arguments supporting the bound do not apply to dS maxima \cite{Montero:2019ekk,Montero:2021otb}, but it is interesting that \eq{fl} is satisfied nonetheless. For the KK photon, the content of the FL bound is equivalent to the statement $m_\text{KK}\gtrsim V^{1/2}$ \cite{Montero:2021otb}, which is satisfied in our $dS_4$ solution. For the vectors obtained from $C_3$, we know that the M2/M5-brane charged states saturate the Weak Gravity Conjecture, so $m\sim g M_p$ and \eq{fl} becomes the statement $g/\sqrt{G}\gtrsim H$. In terms of 11-dimensional quantities, we have
\begin{equation}
    \frac{g}{\sqrt{G}}\sim \frac{\text{vol}(\omega_2)}{\ell_{11}^3}\sim \frac{R^2}{\ell_{11}^3}\gg H \,,
\end{equation}
for electric $C_3$ vectors coming from the reduction of $C_3$ with two legs on $\mathcal{F}_7$, and where $\text{vol}(\omega_2)$ is the volume of the 2-cycle corresponding to the $C_3$ vector field under consideration. Therefore, the bound is satisfied in this case. We should also consider magnetic vectors, obtained from dimensional reduction of the dual potential $C_6$ on a five-cycle of $\mathcal{F}_7$. In this case, the electrically charged object is an M5 brane wrapping this same five-cycle, leading to
\begin{equation}
    \frac{g}{\sqrt{G}}\sim \frac{\text{vol}(\omega_5)}{\ell_{11}^6}\sim \frac{R^5}{\ell_{11}^6}\gg H \,,
\end{equation}
so FL is satisfied for these gauge fields as well.

It is important to keep in mind that this solution arises from solving the Einstein equations perturbatively, in terms of a small parameter $\epsilon$ that controls corrections to a zeroth-order flat background---the RFM itself \cite{ValeixoBento:2025yhz}. Both flux and Casimir contributions act as corrections to this flat background, taking the solution perturbatively away from exact flatness and setting the small parameter $\epsilon$ controlling the expansion. 
The average energy density of the 11-dimensional solution, obtained by dividing the 4d vacuum energy by the volume of the internal manifold, is
\begin{equation} 
    \vev{\rho_{11d}} = 6.08\cdot 10^{-6}\, \ell_{11}^{-11} \,,
\end{equation}
which arises as a balance between the flux and Casimir terms, 
\begin{equation}
    \vev{\rho_{G_4}}\approx 3.65\cdot10^{-5}\, \ell_{11}^{-11} \,,
    \quad \vev{\rho_{\text{Cas}}}\approx -3.04\cdot10^{-5} \ell_{11}^{-11} \,.
    \label{avbar}
\end{equation}
Therefore, the leading-order corrections are controlled by $\epsilon\sim 10^{-5}$. More precisely, any higher-derivative correction to the action with schematic form 
\begin{equation}
    \int P(R_{\mu\nu\rho\sigma}, \vert G_4\vert) \,, 
    \label{hder} 
\end{equation}
where $P$ is some polynomial, will be suppressed by powers of $10^{-5}$. While we do not know the coefficients in front of these higher-derivative corrections (save for a few cases \cite{Hyakutake:2005rb,Hyakutake:2006aq,Hyakutake:2007sm,Grimm:2017okk,Grimm:2017pid}), we expect them to become smaller and smaller as the derivative order gets larger; in fact, assuming a cutoff $\Lambda$ for 11d supergravity that would naturally be expected to appear at or around the 11d Planck scale, the results of \cite{Caron-Huot:2022ugt} would imply that coefficients of higher-derivative operators cannot be more than $\mathcal{O}(1)$ in units of the cutoff to ensure unitarity of the S-matrix (although, strictly speaking, we are outside the regime of validity of this result, since M-theory is not weakly coupled at the cutoff scale); consequently, the solution we found seems quite safe under na\"ive higher-derivative and classical corrections, at least at the homogeneous level. In this regard, it is quite important that the RFM does not contain a non-trivial tadpole equation that would force higher-derivative terms to be a relevant part of the classical solution, avoiding an important subtlety present in standard flux compactifications \cite{Sethi:2017phn}.

While the flux term yields a homogeneous 11-dimensional energy density to leading order, the Casimir energy density coalesces around the locations of the Casimir branes \cite{ValeixoBento:2025yhz}---we therefore expect a Casimir energy profile that is homogeneous along the base $S^1$, but inhomogeneous on the fibre $T^6$.
If the internal profile on $\mathcal{F}_7$ has very steep gradients, so that some regions have energy densities much larger than the average over the RFM, these can show up in corrections like \eq{hder} and invalidate the analysis. This could affect our estimate of $\epsilon$, which is, roughly speaking, the maximum value of the energy density on the internal space, rather than the average. One can see, by looking at the plots in Fig.~\ref{fig:Casimir-branes-Z8}, that the Casimir energy density does not vary significantly over the compact space, despite the presence of Casimir branes, supporting the conclusion that the average energy density is a good estimate for the size of the backreaction. 
Since the Casimir energy density at a maximum, where a Casimir brane is located, is $\hat{\rho}(0,0,0,0,0,0,0)\approx 8.4\cdot 10^{-5}\,\ell_{11}^{-11}$, our estimate of $\epsilon$ from the average $\rho_{\rm Cas}$ is not significantly affected\footnote{This is something that we cannot do conclusively for some supersymmetric solutions, such as DGKT \cite{DeWolfe:2005uu}, due to the presence of singularities in the internal space. The fact that Casimir branes are smooth and have a core profile that can be determined explicitly allows us to avoid all these complications.}.

Since our solution strongly relies on a significant Casimir coefficient, one might wonder whether higher-order loops can be even larger and destroy it. On dimensional grounds, the loop expansion of the Casimir energy (in  11d Planck units) takes the form
\begin{equation}
    V_{\text{Cas}} \sim\frac{1}{R^4}\left[ \mathcal{C}+ \sum_{l=1}^\infty \mathcal{C}_l \left(\frac{\ell_{11}}{R}\right)^{9l}\right] \,,
    \label{Casex}
\end{equation}
with the power of $\ell_{11}$ fixed by the fact that the loop expansion in M-theory is controlled by powers of Newton's constant $G\sim \ell_{11}^9$. 
Unlike the leading-order coefficient $\mathcal{C}$, the higher-order Casimir coefficients $\mathcal{C}_l$ come from divergent loop integrals requiring regularisation, since at higher loops there are unprotected higher-derivative terms in the M-theory action that can contribute to the vacuum energy. 
However, the overall contribution of the Casimir energy is very small in 11-dimensional Planck units because $R\sim 5.16\, \ell_{11}$, which is why the Hubble radius is as large as \eq{hradius}---this was necessary in order to argue stability against classical corrections; this also means that the loop expansion parameter in  \eq{Casex} is
\begin{equation} 
    \left(\frac{\ell_{11}}{R}\right)^{9}\sim 4\cdot 10^{-7} \,,
    \label{lp}
\end{equation}
slightly smaller than the parameter $\epsilon$ controlling classical corrections.  In order to overcome the classical contribution $\mathcal{C}\sim 10^2$ to the Casimir energy, the higher Casimir coefficients would need to satisfy
\begin{equation} 
    \mathcal{C}_l \gtrsim 10^{2+7l} \,.
    \label{rer3r}
\end{equation}
Even if the $\mathcal{C}_l$ are enhanced in the same way that $\mathcal{C}$ was, they still have to beat a factor  \eq{lp} of loop suppression to significantly affect the solution; since $\rho_{11d}$ is small in Planck units, the configuration is quite classical, and quantum corrections are expected to be even smaller. It would nevertheless be interesting to confirm this with an estimate of the coefficients $\mathcal{C}_l$, in terms of explicit expressions that are sums over the KK spectrum.

An important point is that our solution is a perturbative expansion around  a static, classical solution of the equations of motion (the RFM), so that we expect corrections to be convergent when computed around the true static vacuum. This contrasts with what happens in more general, time-dependent situations \cite{Callan:1986bc,Angelantonj:2002ct,Dudas:2004nd,Raucci:2024fnp,Sethi:2017phn,Leone:2025mwo}. For instance, in a non-supersymmetric model with open strings, the one-loop vacuum energy diverges when computed at constant dilaton \cite{Raucci:2024fnp,Leone:2025mwo,Polchinskiv2}, which simply reflects the fact that a constant $\phi$ is not a solution of the classical equations of motion. Since the RFM is a solution to the classical equations of motion, the one-loop calculation is finite. A computation of the higher $\mathcal{C}_l$ would have to be performed through an iterative process where the $(n+1)$--loop calculation is performed around an $n$--loop corrected vacuum; for instance, the two-loop Casimir energy will diverge unless it is computed at the precise point of moduli space where a one-loop maximum is found. While they certainly complicate a detailed treatment, these are all standard issues and do not seem to be a fundamental obstacle to finding a corrected solution.

The analysis up to this point tells us that our $dS_4$ solution is safe from loop and higher-derivative corrections, but one might also worry about non-perturbative effects.
In M-theory, Euclidean M2 and M5 branes wrapping cycles of $\mathcal{F}_7$ lead to well-known non-perturbative effects, whose size can be estimated as
\begin{align} 
    e^{-S_{\text{M2}}}\sim e^{-T_{\text{M2}} \big(\frac18\,R^3\,e^{-2(u+w)}\big)}\approx\, e^{-107} \,,
    && e^{-S_{\text{M5}}}\sim e^{-T_{\text{M5}}\big(R^6\,e^{3u}\big)}\approx e^{-5.54\cdot10^3} \,,
\end{align}
being completely negligible. 
For non-supersymmetric solutions, such as ours, other non-perturbative effects may become important, such as gravitational instantons or bubbles of nothing. On dimensional grounds, the action of any instanton that can be described by 11d supergravity will scale as $(R/\ell_{11})^{9}\sim 10^6$, and would therefore be even more suppressed than the M2/M5 branes described above. 

Despite the fact that our solution appears to be stable against higher-derivative, classical, and quantum corrections, can one really trust a solution whose smallest physical closed curve is just $\frac18 R e^{-3u}\sim 2.54\, \ell_{11}$? Note that our analysis so far has pushed 11-dimensional supergravity all the way down to a few Planck lengths, but we do not know if this is correct; e.g. there might be additional states that are not controlled by supergravity appearing at a mass scale $\mu\lesssim M_{11}$. Such states, if they exist, are necessarily non-supersymmetric; if $\mu$ is close to $M_{11}$, they could be thought of as ``small'' quantum-mechanical black holes and we do expect such objects to exist in the theory, on general grounds. In at least one class of examples---that of AdS$_3$ quantum gravity, where there is more control---we know that there are BTZ black hole states with $\mu=1/(8G)$, and every quantum gravity in AdS$_3$ must have at least one operator other than the graviton with $\mu\leq 3/(32\, G)$ \cite{Benjamin:2019stq}. 
To estimate the effect of such particles, we will pretend the compactification is on a $T^7$ factorised as $T^2\times T^4\times S^1$ with radii $R\,e^{\frac{u}{2} - w}$, $R\,e^{\frac12(u+w)}$ and $\frac18\,R\,e^{-3u}$, respectively. 
The Casimir energy of a particle of mass $\mu$ contributes to the 4d vacuum energy as \cite{Arkani-Hamed:2007ryu}
\begin{equation}
    \delta V^{(4d)}= \left(\frac{8}{R^{7}}\right)^2\,R^7 \left[\sum _{\vec{l}\in\Lambda} \frac{\left(2 \mu ^{11}\right) K_{\frac{11}{2}}(\mu |\vec{l}|)}{(2 \pi )^{\frac{11}{2}} (\mu |\vec{l}|)^{\frac{11}{2}}}\right] \,,
    \label{e343f}
\end{equation}
with the sum running over the defining lattice of the $T^7$.
For our solution, whose smallest cycle is on the $S^1$ base with size $\frac18\,R\,e^{-3u}\sim 2.54\,\ell_{11}$, if $\mu =1$ we have $\delta V^{(4d)}\sim 10^{-9}$, which is similar to the potential at the saddle.
However, this estimate neglects the fact that the spin structure is periodic along this circle; because of this, we expect approximate Bose-Fermi cancellation even at the level of M-theory Planckian black hole microstates. Under these circumstances, it is more appropriate to first reduce to 10d on the supersymmetric circle, and consider the contribution of 11d black holes after compactifying on a non-supersymmetric $T^6$, which amounts (up to $\mathcal{O}(1)$ factors) to using $R\, e^{\frac{u}{2}-w}$ (the second smallest radius) rather than $\frac18\,R\,e^{-3u}$ in estimating \eqref{e343f}. In this case, the contribution of a $\mu\sim1$ black hole is $\delta V^{(4d)}\sim 10^{-12}$; that is, we would need $O(10^2)$ such black hole microstates to affect our solution.

The comments above also apply to string states that may arise from wrapping an M2 brane along the small circle of radius $\frac18 R\,e^{-3u}$. If the radius were sub-Planckian, we would be in the type IIA regime, and the wound M2 states would give us perturbative string states.  Since our radius is slightly super-Planckian, we remain in the M-theory regime, and the states are heavy; the associated mass scale is 
\begin{equation} 
    \sqrt{T}\sim \left[(2\pi^2)^{1/3} \frac18\,R\,e^{-3u}\right]^{1/2}\approx 2.62 \,\ell_{11}^{-1} \,,
\end{equation}
so the wound M2 states are slightly heavier than the black hole states we considered above, and are therefore more suppressed. In fact, since the string is so heavy, these string states will automatically be within the Schwarzschild radius, and therefore, they are part of the black hole spectrum we estimated.

Among the many subtleties involved in estimating the effect of these Planckian states are the effect of including interactions, the lifetime of these states---they would not contribute if they are too short-lived---and even the na\"ive assumption that one is even allowed to run black holes in loops. There are also other possibilities that we are simply unable to properly quantify. 
In summary, even though our estimates suggest that the solution is under control, there is currently no way to control corrections such as those in equation \eq{e343f}, since the spectrum of metastable Planckian states in M-theory is not known.

\subsection{The \texorpdfstring{$dS_4\times T^7/\Z_8$}{dS5xT7/Z8} vacuum of M-theory: An executive summary}

We have found a $dS_4\times T^7/\Z_8$ solution of 11-dimensional supergravity, with an internal compact space that is a quotient of $T^7$ by a fixed-point-free $\mathbb{Z}_8$ isometry breaking all supersymmetries. A classical saddle point is achieved through the balance between Casimir energies and $G_4$ flux. The vacuum energy is small in Planck units, the masses of tachyons and other light fields are of the order of the Hubble scale, the solution is scale separated, and it is protected against all known higher-derivative, loop, and classical corrections. Some important features of the solution are summarised in Table \ref{resumen} for rapid reference.

\begin{landscape}
\renewcommand{\arraystretch}{1.4}
\begin{table}
\centering
\begin{tabular}{ccc}
\hline\textbf{Property}&\textbf{Value}&\textbf{Comments}\\\hline\hline
4d vacuum energy & $V^{(4d)}\approx5.02\cdot 10^{-10}\MPl^{4}$& \begin{tabular}{@{}c@{}} Quite small due to large\\ Casimir coefficient $\mathcal{C}$ and Weyl rescaling\end{tabular} \\\hline
Hubble radius& $H_0^{-1}\sim 7.73\cdot 10^4\, \ell_4 = 7.03\cdot 10^2\, \ell_{11}$&\begin{tabular}{@{}c@{}}$\ell_{11}/\ell_{4}\sim 110$\\ (species vs. 4d Planck scales)\end{tabular}\\\hline
Internal space volume& $\mathcal{V}= 1.21\cdot 10^4\,\ell_{11}^7=2.36\cdot10^{18}\,\ell_4^7$& Large volume helps with control \\\hline
Internal manifold radii 
&\begin{tabular}{@{}r@{}l@{}}
    $R\,e^{\frac{u}{2}-w}$ & $~\sim3.94\, \ell_{11}=434\, \ell_4$ \,, \\ $R\,e^{\frac12(u+w)}$ & $~\sim4.18\, \ell_{11}=460\, \ell_4$ \,, \\
    $\frac18 R\,e^{-3u}$ & $~\sim2.54\,\ell_{11}=279\, \ell_4$
\end{tabular}
&\begin{tabular}{@{}c@{}} Potential impact of black\\ hole-like states due to small $R_i$\end{tabular} \\\hline
Scale separation& Yes: $ \frac18 R\,e^{-3u}\sim 3.61\cdot10^{-3}\, H_0^{-1}$ & Facilitates 4d EFT validity\\\hline
Geometric moduli& \begin{tabular}{@{}c@{}}Four, with masses $m_i^2=$ \\ $\{-35.3\,,\,-12\,,\,-323\,,\,109.8\}\cdot H_0^2$ \end{tabular}&\begin{tabular}{@{}c@{}}All masses light, of order $H_0$;\\ complies with \cite{Ooguri:2018wrx}\end{tabular}\\\hline
Axions & Five, with negligible masses &Generated by Euclidean M2/M5\\\hline
Vectors&Four: one KK photon, three from $C_3$ &Three exactly massless; one massive\\\hline
Light fermions& None &\begin{tabular}{@{}c@{}} Due to antiperiodic spin \\ structure on covering $T^7$\end{tabular}\\\hline
\begin{tabular}{@{}c@{}}11d average\\ energy density\end{tabular}& $\vev{\rho_{11d}}=6.08\cdot10^{-6}\ell_{11}^{-11}$&Control parameter for corrections $\epsilon\sim 10^{-5}$ \\\hline
\begin{tabular}{@{}c@{}}Flux  \& Casimir \\ contributions to $\rho_{11d}$\end{tabular} 
&\begin{tabular}{@{}c@{}} 
    $\vev{\rho_{\text{Cas}}}\approx -3.04\cdot10^{-5}\,\ell_{11}^{-11}$ \\ $\vev{\rho_{G_4}}\approx 3.65\cdot10^{-5}\, \ell_{11}^{-11}$
\end{tabular}
&\begin{tabular}{@{}c@{}}Each contribution larger than $\vev{\rho_{11d}}$,\\ they partially cancel out\end{tabular}\\\hline
\end{tabular}
\caption{Summary of some properties of the $dS_4\times T^7/\Z_8$ solution we found. Here $\ell_{11}$ is the 11-dimensional reduced Planck length, defined by $\kappa_{11}^2=\ell_{11}^9$, and $\ell_4$ is the four-dimensional reduced Planck length, analogously defined.}
\label{resumen}
\end{table}
\renewcommand{\arraystretch}{0.9}
\end{landscape}

\noindent\textbf{Acknowledgements:} 
We thank Bernardo Fraiman and Michelangelo Tartaglia for helpful discussions. We gratefully acknowledge the support of the Spanish State Research Agency (Agencia Estatal de Investigaci\'on) through the grants IFT Centro de Excelencia Severo Ochoa CEX2020-001007-S, PID2021-123017NB-I00, Europa Excelencia EUR2024-153547, RYC2022-037545-I, and CEX2025-001574-S, funded by MICIU/AEI/10.13039/501100011033. The work of MA is supported by the fellowship LCF/BQ/DFR25/12000054 from “La Caixa” Foundation (ID 100010434). The research presented in this publication falls under the research line ``Strings and Quantum Gravity''.

\appendix 

\section{Computing the free part of the integer cohomology of an RFM}
\label{app:A}

Among the basic quantities characterising an RFM are the cohomology groups $H^p(\text{RFM},\mathbb{Z})$, and the corresponding Hodge norm of their free part. In this appendix, we generalise the results of \cite{ValeixoBento:2025yhz} and provide an algorithm to determine these from the defining data of a general RFM. 

We will start with the description of the RFM as a quotient $\mathcal{F}_k\equiv T^k/\Gamma$ of a torus $T^k$ by a finite group $\Gamma$. The quotient map $\pi:\, T^k\rightarrow \mathcal{F}_k$ allows one to pull back $p$-forms $\omega_p$ on $\mathcal{F}_k$ to $T^k$, so that $H^p(\mathcal{F}_k,\mathbb{Z})$ can be identified with a sublattice of $H^p(T^k,\mathbb{Z})$\footnote{We ignore torsion in this appendix; all statements are about the free part of the relevant cohomology groups.}. This sublattice is determined by the condition
\begin{equation}
    \int_{\mathcal{C}_p}\pi^*(\omega_p)\in\mathbb{Z} \,,
    \label{pip}
\end{equation}
for every $\omega_p\in H^p(\mathcal{F}_k,\mathbb{Z})$, and for any $p$-chain $\mathcal{C}_p$ with the property that $\pi(\mathcal{C}_p)$ is a $p$-cycle in $\mathcal{F}_k$. To determine the set of such $\mathcal{C}_p$, we proceed as follows: for a given $p$-chain $\mathcal{C}_p$ in $T^k$, we construct another $p$-chain
\begin{equation}
    \mathcal{I}_p\equiv \sum_{\gamma\in\Gamma} \gamma(\mathcal{C}_p) \,,
    \label{re4}
\end{equation}
i.e. $\mathcal{I}_p$ is the linear combination of the images of $\mathcal{C}_p$ under all elements of the group $\Gamma$. If $\partial\mathcal{I}_p\neq0$, then $\partial \pi(\mathcal{I}_p)\neq0$ as well, since $\mathcal{I}_p$ is invariant under $\Gamma$ by construction. Thus, if $\partial \pi(\mathcal{C}_p)=0$, so that $\partial \pi(\mathcal{I}_p)=0$, we must have $\partial\mathcal{I}_p=0$. Conversely, if $\mathcal{I}_p$ is a boundary, then so is $\pi(\mathcal{I}_p)$ by pushforward. 

The conclusion is that for those $\mathcal{C}_p$ such that $\pi(\mathcal{C}_p)$ is a $p$-cycle in $\mathcal{F}_k$, $\mathcal{I}_p$ is a $p$-cycle in $T^k$. The set of all possible $\mathcal{I}_p$ consists of those homology classes in $H_p(T^k,\mathbb{Z})$ which can be represented by submanifolds invariant under $\Gamma$. Furthermore, we have
\begin{equation}
    \int_{\gamma(\mathcal{C}_p)} \pi^*(\omega_p)= \int_{\pi(\mathcal{C}_p)}\omega_p \,,
\end{equation}
since any chain $\mathcal{C}_p$ and $\gamma(\mathcal{C}_p)$ have the same image in $\mathcal{F}_k$. This means that
\begin{equation} 
    \int_{\mathcal{C}_p} \pi^*(\omega_p)=\frac{1}{\text{ord}(\Gamma)}\int_{\mathcal{I}_p}\pi^*(\omega_p) \,,
    \label{gomamon}
\end{equation}
so the condition \eq{pip} simply becomes
\begin{equation}
    \int_{\mathcal{I}_p}\pi^*(\omega_p) \in\,\text{ord}(\Gamma)\, \mathbb{Z} \,,
    \label{key}
\end{equation}
for all $p$-cycles $\mathcal{I}_p$ of $T^k$ that admit $\Gamma$-invariant representatives. 

In general, determining the possible set of $\mathcal{I}_p$ is difficult. However, for a cyclic RFM which is a fibration of the form $T^{k-1}\rightarrow \mathcal{F}_k \rightarrow S^1$, the task is doable. Any $(p-1)$-cycle  $\mathcal{A}_{p-1}$ in $T^{k-1}$ that is invariant under $\Gamma$ gives rise to a $p$-cycle on $\mathcal{F}_k$, by taking the product with the base circle. Since the generator $\D[g]\,\vec{z}+\bvec[g]$ of the quotient has the shift $\bvec[g]$ entirely along the base, $\mathcal{A}_{p-1}\times S^1$ is a valid $\mathcal{I}_p$. The only other independent possibility is a $p$-cycle  $\mathcal{A}_{p}$ along the fibre, which descends to a $p$-cycle in $\mathcal{F}_k$. The corresponding $\mathcal{I}_p$ is obtained by acting on the pullback of the cycle (located, say, at $(\mathcal{A}_{p},\theta_0)$, where $\theta_0$ is a coordinate along $S^1$) with all elements in $\Gamma$. Denoting the action of $\gamma$ on $\mathcal{A}_{p}$ as $\gamma_* \mathcal{A}_p$, we find that $\mathcal{I}_p$ is equivalent to the fibre class $(\mathcal{D}_p,0)$, where
\begin{equation}
    \mathcal{D}_p\equiv \sum_{\gamma\in \Gamma} \gamma_* \mathcal{A}_p \,. 
    \label{Dpdef}
\end{equation}
For a cyclic RFM which is also a $T^{k-1}$ fibration, each of the $\gamma_* \mathcal{A}_p$ is itself an integral cycle, so the period of any invariant form on $\mathcal{D}_p$ equals $\text{ord}(\Gamma)$ times the period of the original form. As a result, no new conditions arise for fibre forms---any fibre form that was already properly quantised on $T^{k-1}$ will work. The only nontrivial conditions imposed by \eq{key} are therefore on $p$-forms of the form $\omega_p=\eta_{p-1}\wedge d\theta$, where the form $\eta_{p-1}$ should be invariant under the action of $\Gamma$, and $\int_{\mathcal{A}_{p-1}} \eta_{p-1}\in \text{ord}(\Gamma)\,\mathbb{Z}$ for every invariant cycle $\mathcal{A}_{p-1}$.

Finally, the Hodge norm in the RFM can be obtained simply by computing the Hodge norm of $\omega_p$ in $T^k$, divided by  the order of $\Gamma\equiv \mathcal{B}/\text{trans}(\mathcal{B})$, to account for the smaller volume of the quotient RFM; see \cite{ValeixoBento:2025yhz} for further results and a review.

We will now apply these ideas to recover the calculation of \cite{ValeixoBento:2025yhz} for $T^6/\mathbb{Z}_8$, and then perform the equivalent calculation for $T^7/\mathbb{Z}_8$ in the main text:
\begin{itemize}
    \item For the $\mathbb{Z}_8$ action in \cite{ValeixoBento:2025yhz}, we will compute the conditions on integral 2-forms. As explained in that reference, there is a single invariant form involving the base, $\omega_3$. Denoting the elementary 2-planes of the torus as $E_{ij}$, the cycle $E_{56}$ is invariant. The condition \eq{key} then implies
    \begin{equation} 
        \int_{E_{56}} c\, \omega_3\in8\mathbb{Z}\quad\Rightarrow\quad c=8k \,,\quad k\in\mathbb{Z} \,.
    \end{equation}
    Since there are no conditions on forms coming from the fibre, we conclude that the pullback lattice is generated by $\omega_1,\omega_2$ and $8\omega_3$, in agreement with \cite{ValeixoBento:2025yhz}. 

    \item We now turn to studying the closely related $T^7/\mathbb{Z}_8$ example of this paper. One important difference is that, as explained in the main text, this time we are after $H^3(T^7/\mathbb{Z}_8,\mathbb{Z})_{\text{free}}$. There are now five invariant 3-forms, which can be found in \eq{invariant2forms}; of these, only $\omega_1,\omega_2$ and $\omega_3$ involve the base. Denoting the elementary two-cycles of $T^7$ as $E_{ij}$ as above, a basis of invariant 2-cycles is given by\footnote{These can be obtained by finding the orbits of the generators and dividing by the order of the orbit.}
\begin{equation} 
    \mathcal{A}_2^{(1)}=E_{34}+E_{36}+E_{45}+E_{56} \,,
    \quad \mathcal{A}_2^{(2)}=E_{35}+E_{46} \,,
    \quad \mathcal{A}_2^{(3)}=E_{12} \,.
\end{equation}
The periods of the forms $\omega_1,\omega_2,\omega_3$ on these cycles can be arranged in the diagonal $3\times 3$ matrix
\begin{equation} 
    \int_{\mathcal{A}_2^{(i)}\times S^1}\omega_j= \left(\begin{array}{*3{C{1em}}}
        4 & 0 & 0 \\
        0 & 2 & 0 \\
        0 & 0 & 1
    \end{array}\right) \,.
\end{equation}
Condition \eq{key} then determines the corresponding generators of the pullback lattice as $2\omega_1,4\omega_2,8\omega_3$. Since there are no conditions on the fibre forms $\omega_4,\omega_5$, a basis of properly quantised 3-forms is given by $\{2\omega_1,4\omega_2,8\omega_3,\omega_4,\omega_5\}$, which is what was used in the main text.
\end{itemize}

\bibliographystyle{utphys}
\bibliography{references}

\end{document}